%% file: ManuscriptV2.tex
\documentclass[twocolumn,american,reprint, floatfix,amsmath,amssymb, superscriptaddress,footinbib,longbibloiography,prx,aps]{revtex4-1}
\usepackage[T1]{fontenc}
\usepackage[latin9]{inputenc}
\usepackage{geometry}
\usepackage{color}
\usepackage{babel}
\usepackage{amsmath}
\usepackage{amssymb}
\usepackage{graphicx}
\usepackage[unicode=true,pdfusetitle,
 bookmarks=true,bookmarksnumbered=false,bookmarksopen=false,
 breaklinks=false,pdfborder={0 0 1},backref=false,colorlinks=true]
 {hyperref}
\hypersetup{
 urlcolor=blue}
\begin{document}
\title{Boundary central charge from bulk odd viscosity - chiral superfluids}
\author{Omri Golan}
\email{golanomri@gmail.com}

\affiliation{Department of Condensed Matter Physics, Weizmann Institute of Science,
Rehovot 76100, Israel}
\author{Carlos Hoyos}
\affiliation{Department of Physics, Universidad de Oviedo, c/ Federico Garcia Lorca
18, 33007, Oviedo, Spain}
\author{Sergej Moroz}
\affiliation{Physik-Department, Technische Universit{\"a}t M{\"u}nchen, D-85748
Garching, Germany}
\affiliation{Munich Center for Quantum Science and Technology (MCQST), Schellingstr. 4,
D-80799 M{\"u}nchen}
\begin{abstract}

We derive a low energy effective field theory for chiral superfluids,
which accounts for both spontaneous symmetry breaking and fermionic
ground-state topology. Using the theory, we show that the odd (or
Hall) viscosity tensor, at small wave-vector, contains a dependence
on the chiral central charge $c$ of the boundary degrees of freedom,
as well as additional non-universal contributions. We identify related
bulk observables which allow for a bulk measurement of $c$. In Galilean
invariant superfluids, only the particle current and density responses
to strain and electromagnetic fields are required. To complement our results, the effective theory is benchmarked against a perturbative computation within a canonical microscopic model.

\end{abstract}
\maketitle
\global\long\def\ket#1{\left|#1\right\rangle }%
\global\long\def\bra#1{\left\langle #1\right|}%
\global\long\def\braket#1#2{\left\langle #1\right|\left.#2\right\rangle }%
\global\long\def\slashed#1{\not\mathrlap{#1}}%

\section{Introduction}

The odd (or Hall) viscosity $\eta_{\text{o}}$ is a non-dissipative,
time reversal odd, stress response to strain-rate \citep{PhysRevLett.75.697,Avron1998,PhysRevB.86.245309,hoyos2014hall,PhysRevE.89.043019},
which can appear even in superfluids (SFs) and incompressible (or
gapped) fluids,  where the more familiar dissipative
viscosity vanishes. Observable signatures of $\eta_{\text{o}}$
are actively studied in a variety of systems \citep{PhysRevE.90.063005,*PhysRevB.94.125427,*PhysRevLett.119.226602,*PhysRevLett.118.226601,*PhysRevB.96.174524,*PhysRevFluids.2.094101,*banerjee2017odd,*bogatskiy2018edge,*holder2019unified,*PhysRevLett.122.128001},
and recently led to its measurement in a colloidal fluid \citep{soni2018free}
and in graphene \citep{Berdyugineaau0685}.

In isotropic 2+1 dimensional fluids, the odd viscosity tensor at
zero wave-vector ($\mathbf{q}=\mathbf{0}$) reduces to a single component.
In analogy with the celebrated quantization of the odd (or Hall) conductivity
in the quantum Hall (QH) effect \citep{thouless1982quantized,*avron1983homotopy,*golterman1993chern,*qi2008topological,*Nobel-2016,*mera2017topological},
this component obeys a quantization condition 
\begin{align}
 & \eta_{\text{o}}^{\left(1\right)}=-\left(\hbar/2\right)s\cdot n_{0},\;s\in\mathbb{Q},\label{eq:1-1}
\end{align}
in incompressible quantum fluids \citep{PhysRevLett.75.697,Read:2009aa,read2011hall}.
Here $n_{0}$ is the ground state density, and $s$ is a rational
topological invariant  labeling the many-body ground state, which
corresponds to the average angular momentum per particle (in
units of $\hbar$, henceforth set to 1).  

\begin{figure}[!th]
\begin{centering}
\par\end{centering}
\begin{centering}
\par\end{centering}
\begin{centering}
\includegraphics[width=1\linewidth]{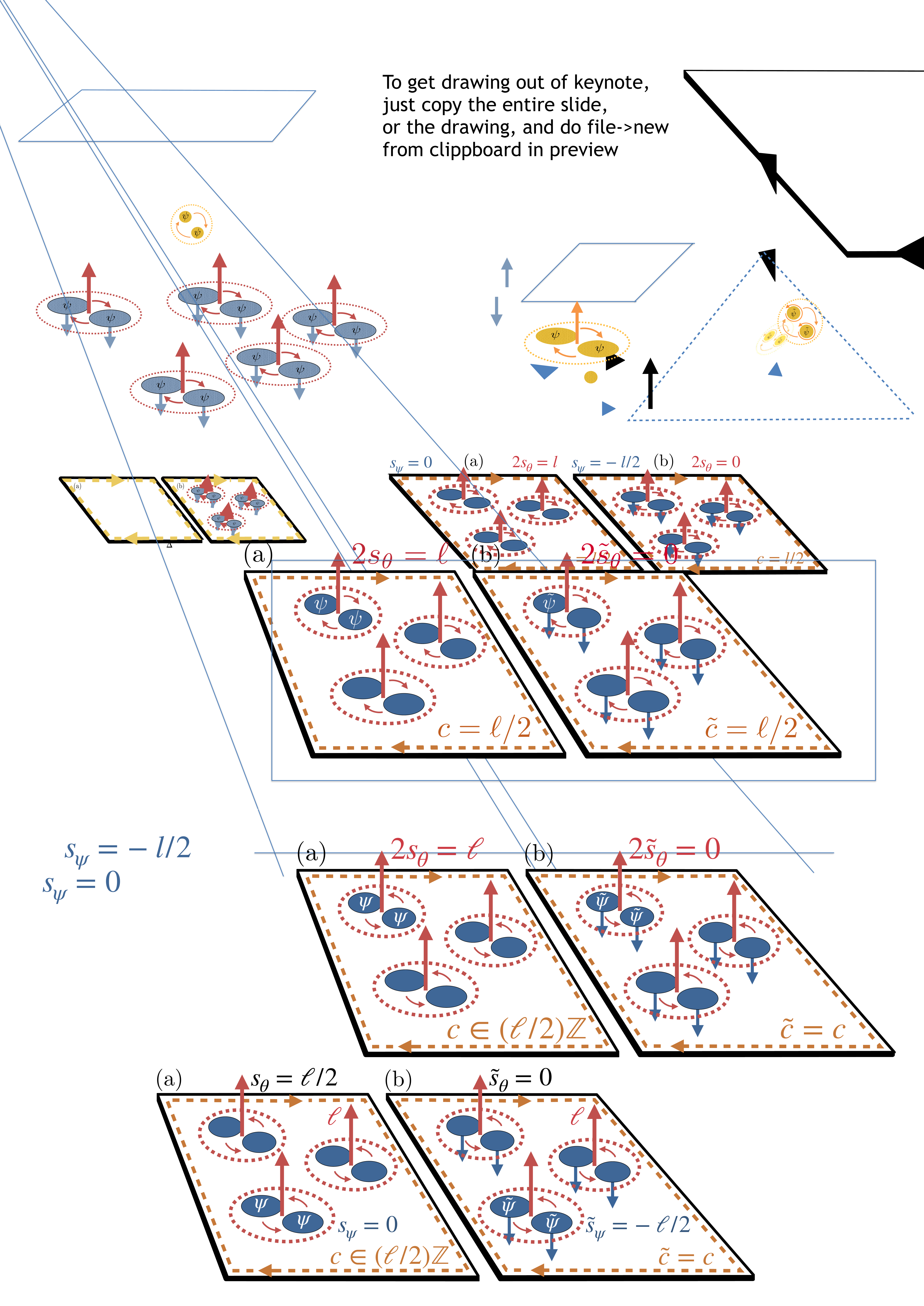}
\par\end{centering}
\caption{(a) A CSF is comprised of fermions $\psi$ which
carry no \textit{geometric} spin, $s_{\psi}=0$, and form Cooper
pairs with a relative angular momentum $\ell\in\mathbb{Z}$ (red arrows). The geometric spin $s_{\theta}=\ell/2$ of the Cooper pair gives rise to the $\mathbf{q}=\mathbf{0}$
odd viscosity \eqref{eq:1-1}, with $s=s_{\theta}$. The CSF
supports boundary degrees of freedom (dashed orange) with a chiral
central charge $c\in\left(\ell/2\right)\mathbb{Z}$, which cannot
be extracted from the odd viscosity $\eta_{\text{o}}\left(\mathbf{q}\right)$
alone \eqref{eq:14}. (b) In an auxiliary CSF, the fermion $\tilde{\psi}$
is assigned a geometric spin $\tilde{s}_{\psi}=-\ell/2$ (blue arrows). The geometric spin of the Cooper pair therefore vanishes,
$\tilde{s}_{\theta}=\ell/2+\tilde{s}_{\psi}=0$, as in an $s$-wave
superfluid, but the central charge is unchanged, $\tilde{c}=c$. As
a result, the small $\mathbf{q}$ behavior of the odd viscosity $\tilde{\eta}_{\text{o}}$ depends only on $c$ \eqref{eq:18-1}. The improved
odd viscosity of the CSF is defined as the odd viscosity of the auxiliary CSF, and is given explicitly by \eqref{eq:16-2}. \label{fig:Comparison-of-the-1-1}}
\end{figure}

Remarkably, Eq.\eqref{eq:1-1} also holds in certain  compressible quantum
fluids, which are the subject of this paper. These are  chiral superfluids
(CSFs), where the ground state is a condensate of Cooper pairs of
fermions, which are spinning around their centre of mass with an angular
momentum $\ell\in\mathbb{Z}$ \citep{read2000paired,volovik2009universe,vollhardt2013superfluid},
see Fig.\ref{fig:Comparison-of-the-1-1}(a). Thin films of $^{3}\text{He-A}$
are experimentally accessible $p$-wave ($\ell=\pm1$) CSFs \citep{Levitin841,*PhysRevLett.109.215301,*Ikegami59,*Zhelev_2017},
and there are proposals for the realization of various  CSFs in cold
atoms \citep{PhysRevLett.101.160401,*PhysRevLett.103.020401,*PhysRevA.86.013639,*PhysRevA.87.053609,*PhysRevLett.115.225301,*BOUDJEMAA20171745,*Hao_2017}.
Closely related chiral superconductors \citep{Kallin_2016,*Sato_2017}
have recently been realized \citep{PhysRevLett.114.017001,*M_nard_2017},
and some of the most debated fractional QH states \citep{PhysRevB.98.045112,*PhysRevLett.121.026801,*PhysRevB.97.121406,*PhysRevB.98.167401,*PhysRevB.99.085309}
are believed to be CSFs of composite fermions \citep{read2000paired,PhysRevX.5.031027,*Son_2018}.
Computing $\eta_{\text{o}}^{\left(1\right)}$ in an $\ell$-wave
CSF, one finds Eq.\eqref{eq:1-1} with the intuitive $s=\ell/2$ \citep{Read:2009aa,read2011hall,hoyos2014effective,shitade2014bulk,moroz2015effective}.
Thus, a measurement of $\eta_{\text{o}}^{\left(1\right)}$ at $\mathbf{q}=\mathbf{0}$
can be used to obtain the angular momentum of the Cooper pair, but
carries no additional information. 

An $\ell$-wave pairing involves the spontaneous symmetry breaking
(SSB) of time reversal $T$ and parity (spatial reflection) $P$ down
to $PT$, and of the symmetry groups generated by particle number
$N$ and angular momentum $L$ down to a diagonal subgroup 
\begin{align}
 & U\left(1\right)_{N}\times SO\left(2\right)_{L}\rightarrow U\left(1\right)_{L-\left(\ell/2\right)N},\label{eq:2-1}
\end{align}
  which implies a single Goldstone field, charged under the broken
generator $N+\left(\ell/2\right)L$, as well as massive Higgs fields
\citep{brusov1981superfluidity,*Volovik_2013,*Sauls:2017aa,*hsiao2018universal,PhysRevB.98.064503}.
For CSFs, it is this SSB pattern, rather than ground-state topology,
which implies the quantization $s=\ell/2$ \citep{hoyos2014effective}
\footnote{As can be seen by considering a mixture of CSFs with different $\ell$s,
where $N$ and $L$ are completely broken, and $s\equiv-2\eta_{\text{o}}^{\left(1\right)}/n=\sum_{i}n_{i}\left(\ell_{i}/2\right)/\sum_{i}n_{i}$
is no longer quantized. }. 

Nevertheless, a CSF with fixed $\ell$ does have a non-trivial ground-state
topology - single fermion excitations  are gapped, and the fermionic
ground state  can be assigned a topological invariant.  This is
the boundary chiral central charge $c\in\left(\ell/2\right)\mathbb{Z}$
(per spin component)  \citep{read2000paired,volovik2009universe},
which counts the net chirality of 1+1 dimensional Majorana spinors
present on the boundary between the CSF and vacuum. For example, a $p$-wave CSF comprised of spin-less fermions has a minimal non-vanishing $c=\pm1/2$, while a $d$-wave ($\ell=\pm2$) CSF, which requires spin-full fermions, has a minimal non vanishing $c=\pm2$, or $c=\pm1$ per spin component.

The invariant
$c$ determines the boundary gravitational anomaly \citep{alvarez1984gravitational,*bertlmann2000anomalies,*bastianelli2006path},
and the\textit{ boundary} thermal Hall conductance  \citep{kane1997quantized,read2000paired,cappelli2002thermal},
 which has been measured in recent experiments on QH and spin systems
\citep{jezouin2013quantum,*banerjee2017observed,*banerjee2018observation,*Kasahara:2018aa}.
Based on the fundamental principle of anomaly inflow \citep{kraus2006holographic,ryu2012electromagnetic,stone2012gravitational,RevModPhys.88.035001}
it is expected that $c$ can be measured in the \textit{bulk} of a
CSF, but whether this is indeed the case, and if so, what should actually
be measured, has so far remained unclear. Providing an answer to this
question is the main goal of the present paper. 

 Analysis of the problem has previously been carried out only within
the \textit{relativistic limit} of the $p$-wave CSF, where the non-relativistic
kinetic energy of the fermions is neglected \citep{volovik1990gravitational,read2000paired,wang2011topological,ryu2012electromagnetic,bradlyn2015topological,PhysRevB.98.064503}.
Within this limit one finds a bulk gravitational Chern-Simons (gCS)
term, which implies a $c$-dependent correction to $\eta_{\text{o}}^{\left(1\right)}$
of \eqref{eq:1-1} at small non-zero wave-vector \citep{abanov2014electromagnetic,klevtsov2015geometric,bradlyn2015topological},
\begin{align}
 & \delta\eta_{\text{o}}^{\left(1\right)}\left(\mathbf{q}\right)=-\frac{c}{24}\frac{1}{4\pi}q^{2}.\label{eq:2}
\end{align}
One is therefore led to suspect that $c$ can be obtained from the
$q^{2}$ correction to $\eta_{\text{o}}$,  but the fate of this correction
beyond the relativistic limit remains unclear.

In particular, the relativistic limit misses most of the physics of
the Goldstone field \citep{PhysRevB.98.064503}. Analysis of the Goldstone
physics in CSFs was undertaken in \citep{volovik1988quantized,*goryo1998abelian,*goryo1999observation,*furusaki2001spontaneous,*stone2004edge,*roy2008collective,*lutchyn2008gauge,ariad2015effective}.
More recently, Refs.\citep{hoyos2014effective,moroz2015effective}
considered CSFs in curved (or strained) space, following the pioneering
work \citep{son2006general} on $s$-wave ($\ell=0$) SFs. These works
demonstrated that the Goldstone field, owing to its charge $L+\left(\ell/2\right)N$,
produces the $\mathbf{q}=\mathbf{0}$ odd viscosity \eqref{eq:1-1},
and it is therefore natural to expect that a $q^{2}$ correction similar
to \eqref{eq:2} will also be produced. Nevertheless, Refs.\citep{hoyos2014effective,moroz2015effective}
did not consider the derivative expansion to the high order at which
$q^{2}$ corrections to $\eta_{\text{o}}$ would appear, nor did
they detect any bulk signature of $c$ at lower orders. 

In this paper we obtain a low energy effective field theory that
captures both SSB and fermionic ground state topology, which extends
and unifies the aforementioned results of \citep{son2006general,hoyos2014effective,moroz2015effective}
and \citep{volovik1990gravitational,read2000paired,wang2011topological,ryu2012electromagnetic,bradlyn2015topological,PhysRevB.98.064503}.
Using the theory we compute the $q^{2}$ correction to $\eta_{\text{o}}$,
and provide several routes towards the bulk measurement of the boundary
central charge in CSFs.

We note that there is an ongoing discussion in the literature regarding
a possible \textit{bulk} thermal Hall conductivity proportional
to $c$, including some contradicting results \citep{qin2011energy,*shitade2014heat,*PhysRevLett.114.016802,*nakai2016finite,*nakai2017laughlin,*kapustin2019thermal,bradlyn2015low}.
This provides further motivation to study the appearance of $c$ in the bulk odd viscosity.

\textcolor{red}{}

\textcolor{red}{}

\section{Building blocks of the effective field theory\label{sec: building blocks} }

In order to probe a CSF, we minimally couple it to two background
fields - a time-dependent spatial metric $G_{ij}$, which we use to
apply strain $u_{ij}=\left(G_{ij}-\delta_{ij}\right)/2$ and strain-rate
$\partial_{t}u_{ij}$, and a $U\left(1\right)_{N}$-gauge field $A_{\mu}=\left(A_{t},A_{i}\right)$,
where we absorb a chemical potential $A_{t}=-\mu+\cdots$. The microscopic
action $S$ is then invariant under $U\left(1\right)_{N}$ gauge transformations,
implying the number conservation $\partial_{\mu}(\sqrt{G}J^{\mu})=0$,
where $\sqrt{G}J^{\mu}=-\delta S/\delta A_{\mu}$. It is also clear
that $S$ is invariant under \textit{spatial} diffeomorphisms generated
by $\delta x^{i}=\xi^{i}\left(\mathbf{x}\right)$, if $G_{ij}$ transforms
as a tensor and $A_{\mu}$ as a 1-form.  Less obvious is the fact
that a Galilean invariant fluid is additionally symmetric under $\delta x^{i}=\xi^{i}\left(t,\mathbf{x}\right)$,
provided one adds to the transformation rule of $A_{i}$ a non-standard
mass-dependent piece \citep{son2006general,hoyos2012hall,hoyos2014effective,gromov2014density,Andreev:2014aa,geracie2015spacetime,Andreev:2015aa,Geracie:2017aa},
\begin{align}
 & \delta A_{i}=-\xi^{k}\partial_{k}A_{i}-A_{k}\partial_{i}\xi^{k}+mG_{ij}\partial_{t}\xi^{j}.\label{eq:4-3-1-1}
\end{align}
We refer to $\delta x^{i}=\xi^{i}\left(\mathbf{x},t\right)$ as \textit{local
Galilean symmetry} (LGS), as it can be viewed as a local version of
the Galilean transformation $\delta x^{i}=v^{i}t$. The LGS implies
the momentum conservation law 
\begin{align}
 & \frac{1}{\sqrt{G}}\partial_{t}\left(\sqrt{G}mJ^{i}\right)+\nabla_{j}T^{ji}=nE_{i}+\varepsilon^{ij}J_{j}B,\label{eq:5-3-1}
\end{align}
where $\sqrt{G}T^{ij}=2\delta S/\delta G^{ij}$ is the stress tensor
and the right hand side is the Lorentz force. This fixes the momentum
density $P^{i}=mJ^{i}$ - a familiar Galilean relation.

Since CSFs spontaneously break the rotation symmetry in flat space,  in order to describe them in curved, or strained, space, it is necessary to introduce a background vielbein. This is
a field $E_{\;j}^{A}$ valued in $GL\left(2\right)$, such that $G_{ij}=E_{\;i}^{A}\delta_{AB}E_{\;j}^{B}$, where 
$A,B\in\left\{ 1,2\right\} $. For a given metric $G$ the vielbein
$E$ is not unique - there an internal $O\left(2\right)_{P,L}=\mathbb{Z}_{2,P}\ltimes SO\left(2\right)_{L}$
ambiguity, or symmetry, acting by $E_{\;j}^{A}\mapsto O_{\;B}^{A}E_{\;j}^{B}$,
$O\in O\left(2\right)_{P,L}$. The generators $L,P$ correspond to
\textit{internal} spatial rotations and reflections, and are analogs
of angular momentum and spatial reflection (parity) on the tangent
space. The inverse vielbein $E_{B}^{\;j}$ is defined by $E_{\;j}^{A}E_{B}^{\;j}=\delta_{B}^{A}$.

 The charge $N+\left(\ell/2\right)L$ of the Goldstone field $\theta$
implies the covariant derivative
\begin{align}
 & \nabla_{\mu}\theta=\partial_{\mu}\theta-A_{\mu}-s_{\theta}\omega_{\mu},\label{eq:6}
\end{align}
with a \textit{geometric} spin $s_{\theta}=\ell/2$. Here $\omega_{\mu}$
is the non-relativistic spin connection, an $SO\left(2\right)_{L}$-gauge
field which is $E^{A}_{\; j}$-compatible, see Appendix \ref{sec:Geometric quantities}.  
So far we assumed that the microscopic fermion $\psi$ does not carry
a geometric spin, $s_{\psi}=0$, which defines the physical system
of interest. It will be useful, however, to generalize to $s_{\psi}\in\left(1/2\right)\mathbb{Z}$, where the covariant derivative of the fermion is 
\begin{align}
\nabla_{\mu}\psi=\left(\partial_{\mu}+iA_{\mu}+is_{\psi}\omega_{\mu}\right)\psi.   
\end{align}
A non-zero $s_{\psi}$ modifies the geometric spin of $\theta$ to $s_{\theta}=s_{\psi}+\ell/2$, and the unbroken generator in \eqref{eq:2-1} to $L-s_{\theta}N$. In the special case
$s_{\psi}=-\ell/2$ the Cooper pair is geometrically spin-less and
$L$ is unbroken, as in an $s$-wave SF, see Fig.\ref{fig:Comparison-of-the-1-1}(b).
This $s_{\theta}=0$ CSF is, however, distinct from a conventional
$s$-wave SF, because $P$ and $T$ are still broken down to $PT$,
and we therefore refer to it as a \textit{geometric} $s$-wave (g$s$-wave)
CSF, to distinguish the two. In particular, a central charge $c\neq0$,
which is $P,T$-odd, is not forbidden, and is in fact
independent of $s_{\psi}$. This makes the g$s$-wave CSF particularly
useful for our purposes.

We note that $\omega_{\mu}$ transforms as a 1-form under LGS only
if $B/2m$ is added to $\omega_{t}$ \citep{hoyos2014effective,moroz2015effective},
which we do implicitly throughout the paper. For $\psi$, this is equivalent
to adding  a g-factor $g_{\psi}=2s_{\psi}$ \citep{geracie2015spacetime}.

\section{Effective field theory\label{sec: effective field theory}}

 Based on the above characterization of CSFs, the low energy, long
wave-length, behavior of the system can be captured by an effective
action $S_{\text{eff}}\left[\theta;A,G\right]$, obtained by
integrating out all massive degrees of freedom - the single fermion
excitations and the Higgs fields. In this section we describe a general expression for $S_{\text{eff}}$, compatible
with the symmetries, SSB pattern, and ground state topology of CSFs.

The effective action can be written order by order
in a derivative expansion, with the power counting scheme \citep{son2006general,hoyos2014effective}
\begin{align}
\partial_{\mu}=O\left(p\right),\;A_{\mu},G_{ij}=O\left(1\right),\;\theta=O\left(p^{-1}\right).\label{eq: power count}
\end{align}
The spin connection
is a functional of $G_{ij}$ that involves a single derivative (see Eq.\eqref{eq:49}), 
so $\omega_{\mu}=O\left(p\right)$. Denoting by $\mathcal{L}_{n}$
the term in the Lagrangian which is $O\left(p^{n}\right)$ and invariant
under all symmetries, we have $S_{\text{eff}}=\sum_{n=0}^{\infty}\int\text{d}^{2}x\text{d}t\sqrt{G}\mathcal{L}_{n}$.
The desired $q^{2}$ corrections to $\eta_{\text{o}}$ are $O\left(p^{3}\right)$,
which poses the main technical difficulty. 

The leading order Lagrangian
\begin{align}
 & \mathcal{L}_{0}=P\left(X\right),\;X=\nabla_{t}\theta-\frac{1}{2m}G^{ij}\nabla_{i}\theta\nabla_{j}\theta,\label{eq:9}
\end{align}
was studied in \citep{hoyos2014effective}, and contains
the earlier results of \citep{volovik1988quantized,goryo1998abelian,goryo1999observation,furusaki2001spontaneous,stone2004edge,roy2008collective,lutchyn2008gauge}.
Here $X$ is the unique $O\left(1\right)$ scalar, which reduces
to the chemical potential $\mu$ in the ground state(s) $\partial_{\mu}\theta=0$,
and $P$ is an arbitrary function of $X$  that physically corresponds
to the ground state pressure $P_{0}=P\left(\mu\right)$. The function
$P$ also determines the ground state density $n_{0}=P'\left(\mu\right)$,
and the leading dispersion of the Goldstone mode $\omega^{2}=c_{s}^{2}q^{2}$,
where $c_{s}^{2}=\partial_{n_{0}}P_{0}/m=P'/P''m$ is the speed of
sound, squared. For $\ell\neq0$, the spin connection appears in each $\nabla\theta$
\eqref{eq:6}, and so $\mathcal{L}_{0}$ includes $O\left(p\right)$
contributions, which produce the leading odd viscosity and conductivity,
discussed below. There are no additional terms at $O\left(p\right)$,
so that $\mathcal{L}_{1}=0$ \citep{hoyos2014effective}. 

At $O\left(p^{2}\right)$ one has  
\begin{align}
\mathcal{L}_{2}= & F_{1}\left(X\right)R\label{eq:10-1}\\
 & +F_{2}\left(X\right)\left[mK_{\;i}^{i}-\nabla^{2}\theta\right]^{2}\nonumber \\
 & +F_{3}\left(X\right)\left[2m\left(\nabla_{i}K_{\;j}^{j}-\nabla^{j}K_{ji}\right)\nabla^{i}\theta\right]+\cdots,\nonumber 
\end{align}
where $K_{ij}=\partial_{t}G_{ij}/2$ and $R$ are the extrinsic
curvature and Ricci scalar of the spatial slice at time $t$ \citep{carroll2004spacetime},
the $F$s are arbitrary functions of $X$, and dots indicate additional
terms which do not contribute to $\eta_{\text{o}}$ up to $O\left(p^{2}\right)$, see Appendix \ref{subsec: Second order effective action} for the full expression. 
The Lagrangian $\mathcal{L}_{2}$ was obtained
in \citep{son2006general} for $s$-wave SFs. For $\ell\neq0$ the
spin connection in $\nabla\theta$ produces $O\left(p^{3}\right)$
contributions to $\mathcal{L}_{2}$, and, in turn, non-universal $q^{2}$
corrections to $\eta_{\text{o}}$.

The term $\mathcal{L}_{3}$ is the last ingredient required for reliable
results at $O\left(p^{3}\right)$. Most importantly, it includes the
(non-relativistic) gCS term  \citep{Chern-Simons,jackiw2003chern,kraus2006holographic,witten2007three,perez2010conserved,stone2012gravitational,bradlyn2015topological,gromov2016boundary}
 
\begin{align}
 & \mathcal{L}_{3}\supset\text{\ensuremath{\mathcal{L}}}_{\text{gCS}}=-\frac{c}{48\pi}\omega\text{d}\omega,\label{eq:10}
\end{align}
 where the $c$-dependence is required to match the boundary gravitational anomaly \citep{kraus2006holographic,stone2012gravitational,PhysRevB.98.064503}, and $\omega\text{d}\omega=\varepsilon^{\mu\nu\rho}\omega_{\mu}\partial_{\nu}\omega_{\rho}$.
 Unlike the lower order terms, $\mathcal{L}_{\text{gCS}}$ is independent
of $\theta$, and encodes only the response of the gapped fermions to the background fields.
      In Appendices \ref{subsec:Odd-viscosity-from} and \ref{subsec:Additional terms at third order} we argue that additional terms in $\mathcal{L}_{3}$ do
not produce $q^{2}$ corrections to $\eta_{\text{o}}$. 

There are three topological terms that can be added
to $S_{\text{eff}}$ \citep{ferrari2014fqhe,can2014fractional,abanov2014electromagnetic,gromov2014density,gromov2015framing,can2015geometry,klevtsov2015geometric,bradlyn2015topological,Klevtsov_2016,klevtsov2017laughlin,Cappelli_2018}.
These are the $U\left(1\right)$ Chern-Simons (CS) and first and second
Wen-Zee (WZ1, WZ2) terms, which can be added to $\mathcal{L}_{1}$,
$\mathcal{L}_{2}$, $\mathcal{L}_{3}$, respectively \footnote{These need to be modified as in \citep{hoyos2012hall,Andreev:2014aa,Andreev:2015aa} for LGS.} , 
\begin{align}
 & \frac{\nu}{4\pi}\left(A\text{d}A-2\overline{s}\omega\text{d}A+\overline{s^{2}}\omega\text{d}\omega\right).\label{eq:11}
\end{align}
As our notation suggests, WZ2 and gCS are identical for the purpose
of local bulk responses, of interest here, but the two are globally
distinct \citep{bradlyn2015topological,gromov2016boundary,Cappelli_2018}.
Based on symmetry, and ignoring boundary physics, the independent
coefficients $\nu$, $\nu\overline{s}, \nu\overline{s^{2}}$
obey certain quantization conditions \citep{witten2007three}, but
are otherwise unconstrained. The absence of a boundary $U\left(1\right)_{N}$-anomaly
then fixes $\nu=0$ \citep{PhysRevB.98.064503}, but leaves $\nu\overline{s},\nu\overline{s^{2}}$
undetermined \citep{bradlyn2015topological,gromov2016boundary,Cappelli_2018}.
One can argue that a Chern-Simons term can only appear for the unbroken
generator $L-s_{\theta}N$, so that $\nu=0$ implies $\nu\overline{s}=\nu\overline{s^{2}}=0$.
Moreover, in the following section we will see that a perturbative computation within a canonical model
for $\ell=\pm1$ shows that $\nu\overline{s}=\nu\overline{s^{2}}=0$, which applies to any deformation of the model
(which preserves the symmetries, SSB pattern, and single fermion gap),
due to the quantization of $\nu\overline{s},\nu\overline{s^{2}}$.
Accordingly, we set $\nu\overline{s}=\nu\overline{s^{2}}=0$ in the
following.   

\section{Benchmarking the effective theory against a microscopic model}

In this section we take a complementary approach and compute $S_{\text{eff}}$
perturbatively, starting from a canonical microscopic model for a spinless
$p$-wave CSF. The perturbative computation verifies the general expression
in a particular example, and determines the coefficients of topological
terms which are not completely fixed by symmetry. It also gives one
a sense of the behavior of the coefficients of non-topological terms
as a function of microscopic parameters. Here we will outline the computation and describe its results, deferring many technical details to Appendix \ref{sec:microscopic model }. 

The microscopic model is given by 
\begin{align}
S_{\text{m}}=\int\mbox{d}^{2}x & \text{d}t\sqrt{G}\left[\frac{i}{2}\psi^{\dagger}\overleftrightarrow{\nabla_{t}}\psi-\frac{1}{2m}G^{ij}\nabla_{i}\psi^{\dagger}\nabla_{j}\psi\right.\label{eq:3-1-1}\\
 & \left.\vphantom{\frac{1}{2m}}+\left(\frac{1}{2}\Delta^{j}\psi^{\dagger}\nabla_{j}\psi^{\dagger}+h.c\right)-\frac{1}{2\lambda}G_{ij}\Delta^{i*}\Delta^{j}\right],\nonumber 
\end{align}
where $\nabla_{\mu}\psi=\left(\partial_{\mu}+iA_{\mu}\right)\psi$, so $s_\psi=0$. Apart
from the standard non-relativistic kinetic term, the action includes
the simplest attractive two-body interaction \citep{Volovik:1988aa,quelle2016edge},
mediated by the complex vector $\Delta^{i}$, the order parameter,
with coupling constant $\lambda>0$. 

For a given $\Delta^{j}$, the fermion $\psi$ is gapped, unless the
chemical potential $\mu$ or chirality $\ell=\text{sgn}\left(\text{Im}\left(\Delta^{x}\Delta^{y*}\right)\right)$
 are tuned to 0, and forms a fermionic topological phase characterized
by the boundary chiral central charge \citep{read2000paired,volovik2009universe,ryu2010topological}
\begin{align}
 & c=-\left(\ell/2\right)\Theta\left(\mu\right)\in\left\{ 0,\pm1/2\right\} .\label{eq:12-1}
\end{align}

An effective action $S_{\text{eff},\text{m}}\left[\Delta;A,G\right]$
for $\Delta^{j}$ in the background $A_{\mu},G_{ij}$ is then obtained
by integrating over the fermion. The subscript "m" indicates that
this is obtained from the particular microscopic model $S_{\text{m}}$.
Since Eq.\eqref{eq:3-1-1} is quadratic in $\psi,\psi^{\dagger}$,
obtaining $S_{\text{eff},\text{m}}$ is formally straightforward,
and leads to a functional Pfaffian.

To zeroth order in derivatives, the action $S_{\text{eff},\text{m}}$ is given by a potential for $\Delta^{i}$, which is minimized by the $p_{x}\pm ip_{y}$ configurations. In flat space these are given by the familiar $\Delta^{j}\partial_{j}=\Delta_{0}e^{-2i\theta}\left(\partial_{x}\pm i\partial_{y}\right)$.
Here $\Delta_{0}$ is a fixed function of $m,\mu$ and $\lambda$,
determined by the minimization, while the phase $\theta$ and chirality
$\ell=\pm1$ are undetermined. In order to write down the $p_{x}\pm ip_{y}$
configurations in curved space it is necessary to use a background
vielbein \cite{read2000paired,hoyos2014effective,moroz2015effective,moroz2016chiral,quelle2016edge},
\begin{align}
\Delta^{j} & =\Delta_{0}e^{-2i\theta}\left(E_{1}^{\;j}\pm iE_{2}^{\;j}\right).\label{eq:12}
\end{align}

Fluctuations of $\Delta$ away from these configurations correspond
to massive Higgs modes, which should in principle be integrated out
to obtain a low energy action $S_{\text{eff},\text{m}}\left[\theta;A,G\right]$
that can be compared with the general $S_{\text{eff}}$ of the previous
section. We will simply ignore these fluctuations, and obtain $S_{\text{eff},\text{m}}\left[\theta;A,G\right]$
by plugging Eq.\eqref{eq:12} into $S_{\text{eff},\text{m}}\left[\Delta;A,G\right]$.
This will suffice as a derivation of $S_{\text{eff}}$ from a microscopic
model. A proper treatment of the massive Higgs modes will only further
renormalize the coefficients we find, apart from the central charge
$c$.

To practically compare the actions $S_{\text{eff}}$ and $S_{\text{eff},\text{m}}$
we expand them in fields, to second order around $\theta=0,A_{\nu}=-\mu\delta_{\nu}^{t},G_{ij}=\delta_{ij}$,
and in derivatives, to third order, see Appendices \ref{subsec:Effective-action-and} and \ref{sec:microscopic model }. Equating
these two double expansions leads to an overdetermined system of equations
for the phenomenological parameters in $S_{\text{eff}}$ in terms
of the microscopic parameters in $S_{\text{m}}$, with a unique solution.
In particular, we find the dimensionless parameters
\begin{align}
\frac{P''}{m}= & \frac{1}{2\pi}\begin{cases}
1\\
\frac{1}{1+2\kappa}
\end{cases},\hspace{1.4cm}F_{1}'=\frac{1}{96\pi}\begin{cases}
1\\
\frac{3}{1+2\kappa}
\end{cases},\label{eq:c1-1}\\
mF_{2}= & -\frac{1}{128\pi}\begin{cases}
1+2\kappa\\
\frac{1}{1+2\kappa}
\end{cases},\;mF_{3}=\frac{1}{48\pi}\begin{cases}
1+\kappa\\
\frac{1}{1+2\kappa}
\end{cases},\nonumber \\
c= & \begin{cases}
-\ell/2\\
0
\end{cases},\nonumber 
\end{align}
where $\kappa=\left|\mu\right|/m\Delta_{0}^{2}>0$, and the upper and lower values
refer to $\mu>0$ and $\mu<0$ respectively. We note that for $\mu>0$ there is
a single particle Fermi surface with energy $\varepsilon_{F}=\mu$
and wave-vector $k_{F}=\sqrt{2m\mu}$, which for small $\lambda$
will acquire an energy gap $\varepsilon_{\Delta}=\Delta_{0}k_{F}\ll\varepsilon_{F}$.
In this  weak-coupling regime, it is natural to parametrize the
coefficients in \eqref{eq:c1-1} using the small parameter $\varepsilon_{\Delta}/\varepsilon_{F}=\sqrt{2/\kappa}$.

The coefficient $P''$ determines the leading odd (or Hall) conductivity
and has been computed previously in the literature \citep{volovik1988quantized,goryo1998abelian,goryo1999observation,furusaki2001spontaneous,stone2004edge,roy2008collective,lutchyn2008gauge,ariad2015effective},
while $F_{1},F_{2}$ and $F_{3}$, to the best of our knowledge, have
not been computed previously, even for an $s$-wave SF.

Crucially, Eq.\eqref{eq:c1-1} shows that the coefficient $c$ of the bulk gCS
term \eqref{eq:10} matches the known boundary central charge \eqref{eq:12-1}. It follows that there is no WZ2 term in $S_{\text{eff},\text{m}}$, so $\nu\overline{s^{2}}=0$, in accordance with the previous section. We additionally confirm that $\nu=\nu\overline{s}=0$. The direct confirmation of the gCS term and its coefficient
within a non-relativistic microscopic model has been anticipated for
some time \citep{volovik1990gravitational,read2000paired,wang2011topological,ryu2012electromagnetic,PhysRevB.98.064503},
and is the main result of the perturbative computation.

A few additional comments regarding Eq.\eqref{eq:c1-1} are in order:
\begin{enumerate}
\item The seeming quantization of $P''/m$ and $F_{1}'$ for $\mu>0$ is
a non-generic result, as was shown explicitly for $P''/m$ \citep{ariad2015effective}.
\item The free fermion limit $\kappa\rightarrow\infty$, or $\Delta_{0}\rightarrow0$,
of certain coefficients in \eqref{eq:c1-1} diverges for $\mu>0$
but not for $\mu<0$. This signals the breakdown of the gradient expansion
for a gapless Fermi surface, but not for gapped free fermions.
\item   The opposite limit, $\kappa\rightarrow0$, or $m\rightarrow\infty$,
is the relativistic limit mentioned above, in which the fermionic
part of the model reduces to a 2+1 dimensional Majorana spinor with
mass $\mu$ and speed of light $\Delta_{0}$, coupled to Riemann-Cartan
geometry described by $\Delta^{i},\;A_{\mu}$, and in which $S_{\text{eff},\text{m}}$
was already computed \citep{PhysRevB.98.064503,hughes2013torsional}.
Accordingly, the limit $\kappa\rightarrow0$ of \eqref{eq:c1-1} indeed
reproduces the results of \citep{PhysRevB.98.064503,hughes2013torsional}
in a suitable sense, see Appendix \ref{sec:microscopic model }.   
\end{enumerate}

\section{Induced action and linear response\label{sec: induced action and linear response}}

Having derived and benchmarked the effective theory, we are now in a position to obtain linear response functions, in particular the $q^{2}$ corrections to the odd viscosity, and related observables that allow for the bulk measurement of $c$.

By expanding $S_{\text{eff}}$ to second order in the fields $\theta,A_{t}-\mu,A_{i},u_{ij}$,
and performing Gaussian integration over $\theta$, we obtain an induced
action $S_{\text{ind}}\left[A_{\mu},u_{ij}\right]$ that captures
the linear response of CSFs to the background fields, see Appendix \ref{subsec:Obtaining--from} for explicit expressions. 
Taking functional derivatives, one obtains the expectation values $J^{\mu}=G^{-1/2}\delta S_{\text{ind}}/\delta A_{\mu}$,
$T^{ij}=G^{-1/2}\delta S_{\text{ind}}/\delta u_{ij}$ of the current
and stress, and from them the conductivity $\sigma^{ij}=\delta J^{i}/\delta E_{j}$,
the viscosity $\eta^{ij,kl}=\delta T^{ij}/\delta\partial_{t}u_{kl}$,
 and the mixed response function $\kappa^{ij,k}=\delta T^{ij}/\delta E_{k}=\delta J^{k}/\delta\partial_{t}u_{ij}$.
We will also need the static susceptibilities $\chi_{JJ}^{\mu,\nu},\;\chi_{TJ}^{ij,\nu}$,
defined by restricting to time independent $A_{\mu},u_{ij}$, and
computing $\delta J^{\mu}/\delta A_{\nu}$ and $\delta J^{\nu}/\delta u_{ij}$,
respectively. 

Before computing $\eta_{\text{o}}$, it is useful to restrict its
form based on dimensionality and symmetries: space-time translations, spatial rotations, and $PT$. The analysis is performed in Appendices \ref{subsec: B.1}-\ref{subsec: B.4}, and results in the expression 
\begin{align}
\eta_{\text{o}}\left(\omega,\mathbf{q}\right)= & \eta_{\text{o}}^{\left(1\right)}\sigma^{xz}+\eta_{\text{o}}^{\left(2\right)}\left[\left(q_{x}^{2}-q_{y}^{2}\right)\sigma^{0x}-2q_{x}q_{y}\sigma^{0z}\right],\label{eq:3-3-2-1}
\end{align}
written in 
the basis $\sigma^{ab}=2\sigma^{[a}\otimes\sigma^{b]}$ of anti-symmetrized tensor products of the symmetric Pauli matrices 
\citep{Avron1998}. As components of the strain tensor, the matrices
$\sigma^{x},\sigma^{z}$ correspond to shears, while the identity
matrix $\sigma^{0}$ corresponds to a dilatation. The details of the
system are encoded in two independent coefficients $\eta_{\text{o}}^{\left(1\right)},\eta_{\text{o}}^{\left(2\right)}\in\mathbb{C}$,
which are functions of $\omega,q^{2}$. At $\mathbf{q}=\mathbf{0}$ the odd viscosity
tensor reduces to a single component, $\eta_{\text{o}}\left(\omega,\mathbf{0}\right)=\eta_{\text{o}}^{\left(1\right)}\left(\omega\right)\sigma^{xz}$,
as is well known \citep{PhysRevLett.75.697,Avron1998,PhysRevB.86.245309,hoyos2014hall,PhysRevE.89.043019}.
The additional component $\eta_{\text{o}}^{\left(2\right)}$ has not
been discussed much in the literature \citep{abanov2014electromagnetic,hoeller2018second},
and also appears in the presence of (pseudo-)vector anisotropy
\citep{PhysRevB.99.045141,PhysRevB.99.035427}, in which case $\mathbf{q}$
should be replaced by a background (pseudo-)vector $\mathbf{b}$. The expression
\eqref{eq:3-3-2-1} applies at finite temperature, out of equilibrium,
and in the presence of disorder that preserves the symmetries on average.
For clean systems at zero temperature, $\eta_{\text{o}}^{\left(1\right)},\eta_{\text{o}}^{\left(2\right)}$
are both real, even functions of $\omega$. In gapped systems $\eta_{\text{o}}^{\left(1\right)},\eta_{\text{o}}^{\left(2\right)}$ will usually be regular at $\omega=0=q^{2}$, though exceptions to this rule have recently been found \citep{Rose2019}.

For the CSF, we find the $\omega=0$ coefficients 
\begin{align}
\eta_{\text{o}}^{\left(1\right)}\left(q^{2}\right)= & -\frac{1}{2}s_{\theta}n_{0}-\left(\frac{c}{24}\frac{1}{4\pi}+s_{\theta}C^{\left(1\right)}\right)q^{2}+O\left(q^{4}\right),\nonumber \\
\eta_{\text{o}}^{\left(2\right)}\left(q^{2}\right)= & \frac{1}{2}s_{\theta}n_{0}q^{-2}+\left(\frac{c}{24}\frac{1}{4\pi}+s_{\theta}C^{\left(2\right)}\right)+O\left(q^{2}\right),\label{eq:14}
\end{align}
where $C^{\left(1\right)},C^{\left(2\right)}\in\mathbb{R}$ are generically
non-zero, and are given by particular linear combinations of the dimensionless
coefficients $F_{1}'\left(\mu\right),mF_{2}\left(\mu\right)$ and $mF_{3}\left(\mu\right)$ 
defined in \eqref{eq:10-1}, see Appendix \ref{subsec:Obtaining--from} for more details. 

The leading term in $\eta_{\text{o}}^{\left(1\right)}$ is the familiar
Eq.\eqref{eq:1-1}, which also appears in gapped states, while the non-analytic
leading term in $\eta_{\text{o}}^{\left(2\right)}$ occurs because
the superfluid is gapless, and does not appear when $q\rightarrow0$ at $\omega\neq0$ \citep{hoyos2014effective}.
Both leading terms obey the same quantization condition due to SSB, and are independent of $c$.
The sub-leading corrections to both $\eta_{\text{o}}^{\left(1\right)},\eta_{\text{o}}^{\left(2\right)}$
contain the quantized gCS contributions proportional to $c$, but
also the non-universal coefficients $C^{\left(1\right)},C^{\left(2\right)}$.
Thus $c$ cannot be extracted from a measurement of
$\eta_{\text{o}}$ alone. 

Noting that the non-universal sub-leading corrections to $\eta_{\text{o}}$
originate from the geometric spin $s_{\theta}=\ell/2$ of the Goldstone
field, one is naturally led to consider the g$s$-wave CSF, where
$s_{\theta}=0$ and the odd viscosity is, to leading order in $q$,
purely due to $\mathcal{L}_{\text{gCS}}$,
\begin{align}
\tilde{\eta}_{\text{o}}^{\left(1\right)}\left(q^{2}\right)= & -\frac{c}{24}\frac{1}{4\pi}q^{2}+O\left(q^{4}\right),\label{eq:18-1}\\
\tilde{\eta}_{\text{o}}^{\left(2\right)}\left(q^{2}\right)= & \frac{c}{24}\frac{1}{4\pi}+O\left(q^{2}\right).\nonumber 
\end{align}
Here and below we use $O$ and $\tilde{O}$, for the quantity $O$
in the CSF and in the corresponding $g$s-wave CSF, respectively.
Equation \eqref{eq:18-1} follows from \eqref{eq:14} by setting $s_{\theta}=0$,
but can be understood directly from $S_{\text{eff}}$. Indeed, for
the g$s$-wave CSF, $S_{\text{eff}}$ is identical to that of the
conventional $s$-wave SF to $O\left(p^{2}\right)$, but contains
the additional $\mathcal{L}_{\text{gCS}}$ at $O\left(p^{3}\right)$,
which produces \eqref{eq:18-1}. 

 Due to the LGS \eqref{eq:4-3-1-1}-\eqref{eq:5-3-1}, the viscosity
\eqref{eq:18-1} implies also
\begin{align}
 & \tilde{\chi}_{TJ,\text{o}}^{ij,k}=-\frac{i}{m}\frac{c}{48\pi}q_{\perp}^{i\vphantom{j}}q_{\perp}^{j}q_{\bot}^{k\vphantom{j}}+O\left(q^{4}\right),\label{eq:17}
\end{align}
where $q_{\perp}^{i}=\varepsilon^{ij}q_{j}$, and the subscript ``o''
(``e'') refers to the $P,T$-odd (even) part of an object. Thus, a steady $P,T$-odd current
$\tilde{J}_{\text{o}}^{k}=-\frac{1}{m}\frac{c}{96\pi}\partial_{\perp}^{k}R+O\left(q^{4}\right)$
flows perpendicularly to gradients of curvature $R=-2\partial_{\perp}^{i}\partial_{\perp}^{j}u_{ij}$.
We conclude that, in the g$s$-wave CSF, $c$ can be extracted
from a measurement of $\tilde{\eta}_{\text{o}}$, and in the Galilean
invariant case, also from a measurement of the current $\tilde{J}$
in response to strain. 

Though the simple results above do not apply to the physical system
of interest, the CSF, there is a relation between the observables
of the CSF and the corresponding $g$s-wave CSF, which we can utilize.
At the level of induced actions, it is given by 

\begin{align}
    \tilde{S}_{\text{ind}}\left[A_{\mu},u_{ij}\right]=S_{\text{ind}}\left[A_{\mu}-\left(\ell/2\right)\omega_{\mu},u_{ij}\right],
\end{align}
where $\omega_{\mu}$ is expressed through $u_{ij}$ as in Appendix \ref{sec:Geometric quantities}, 
 and by taking functional derivatives one obtains relations between response functions \citep{geracie2015spacetime}. In
particular, 
\begin{align}
\tilde{\eta}_{\text{o}}^{ij,kl}= & \eta_{\text{o}}^{ij,kl}-\frac{\ell}{4}n_{0}\left(\sigma^{xz}\right)^{ij,kl}\label{eq:16-2}\\
 & +\frac{i\ell}{4}\left(\kappa_{\text{e}\vphantom{\bot}}^{ij,(k}q_{\perp}^{l)}-\kappa_{\text{e}\vphantom{\bot}}^{kl,(i}q_{\perp}^{j)}\right)+\frac{\ell^{2}}{16}\sigma_{\text{o}}q_{\perp}^{(i}\varepsilon_{\vphantom{\bot}}^{j)(k}q_{\perp}^{l)},\nonumber 
\end{align}
where the response functions $\eta_{\text{o}},\sigma_{\text{o}},\kappa_{\text{e}}$
depend on $\omega,\mathbf{q}$. \textcolor{red}{} In a Galilean
invariant system one further has 
\begin{align}
\tilde{\chi}_{TJ,\text{o}}^{ij,k}= & \chi_{TJ,\text{o}}^{ij,k}-\frac{\ell}{4m}\chi_{TJ,\text{e}}^{ij,t}iq_{\bot}^{k}\label{eq:21}\\
 & +\frac{\ell}{2}iq_{\bot}^{(i}\chi_{JJ,\text{e}}^{j),k}+\frac{\ell^{2}}{8m}q_{\bot}^{(i}\chi_{JJ,\text{o}}^{j),t}q_{\bot}^{k},\nonumber 
\end{align}
and we note the relations $\chi_{TJ,\text{e}}^{ij,t}=\kappa_{\text{e}}^{ij,k}iq_{k},\;\chi_{JJ,\text{o}}^{j,t}=\sigma_{\text{o}}q_{\bot}^{j},\;\chi_{JJ,\text{e}}^{j,k}=\rho_{\text{e}}q_{\bot}^{j}q_{\bot}^{k}$,
between the above susceptibilities, the response functions $\kappa_{\text{e}},\sigma_{\text{o}}$,
and the London diamagnetic response $\rho_{\text{e}}$.

\section{Discussion\label{sec:discussion}}

Equations \eqref{eq:18-1} and \eqref{eq:16-2} are the main results
of this paper. They rely on the SSB pattern \eqref{eq:2-1}, but
not on Galilean symmetry. Equation \eqref{eq:16-2} expresses $\tilde{\eta}_{\text{o}}$
as a bulk observable of CSFs, which we refer to as the \textit{improved
odd viscosity}. According to \eqref{eq:18-1}, the leading term in
the expansion of $\tilde{\eta}_{\text{o}}\left(0,\mathbf{q}\right)$
around $\mathbf{q}=\mathbf{0}$ is fixed by $c$. Since this leading
term occurs at second order in $\mathbf{q}$, in order to extract
$c$ one needs to measure $\sigma_{\text{o}},\chi_{\text{e}}$, and
$\eta_{\text{o}}$, at zeroth, first, and second order, respectively.
In a Galilean invariant system, \eqref{eq:18-1} and \eqref{eq:16-2}
imply \eqref{eq:17} and \eqref{eq:21} respectively, which, in turn,
show that $c$ can be extracted in an experiment where $U\left(1\right)_{N}$
fields and strain are applied, and the resulting number current and
density are measured. In particular, a measurement of the stress tensor
is not required. Since $U\left(1\right)_{N}$ fields can be applied
in Galilean invariant fluids by tilting and rotating the sample,
we believe that a bulk measurement of the boundary central charge,
through \eqref{eq:17} and \eqref{eq:21}, is within reach of existing
experimental techniques. 

Finally, we comment on the implications of our results to QH physics. The problem of obtaining $c$ from a bulk observable has been previously
studied in QH states, described by \eqref{eq:10}-\eqref{eq:11} \citep{ferrari2014fqhe,can2014fractional,abanov2014electromagnetic,gromov2014density,gromov2015framing,can2015geometry,klevtsov2015geometric,bradlyn2015topological,Klevtsov_2016,gromov2016boundary,klevtsov2017laughlin,Cappelli_2018}.
 It was found that $c$ can only be extracted if $\text{var}s=\overline{s^{2}}-\overline{s}^{2}=0$, in which case the response to strain, at fixed $A_{\mu}-\overline{s}\omega_{\mu}$,
depends purely on $c$ \citep{bradlyn2015topological,gromov2016boundary}. This is a useful theoretical characterization, which seems challenging experimentally in light of the need to maintain the fine tuned relation $A_{\mu}=\overline{s}\omega_{\mu}$ while the strain $u_{ij}$, and therefore $\omega_{\mu}$, vary in time and space.
 The improved odd viscosity \eqref{eq:16-2}, constructed
here, applies also to $\text{var}s=0$ QH states, with $\ell$ replaced
by $-2\overline{s}$, and defines a bulk observable which
is determined by $c$, and whose measurement does not require such fine tuning.

\textcolor{red}{}
\begin{acknowledgments}
The authors thank Daniel Ariad for sharing and explaining the Mathematica
code used in \citep{ariad2015effective}.  We also benefited from
discussions with Ady Stern, Andrey Gromov, Barry Bradlyn, David Mross,
Micha Berkooz, Ryan Thorngren, Semyon Klevtsov, Tobias Holder, and
Vatsal Dwivedi. O.G. acknowledges support from the Deutsche Forschungsgemeinschaft
(DFG, CRC/Transregio 183, EI 519/7-1), the Israel Science Foundation
(ISF), and the European Research Council (ERC, Project LEGOTOP). C.H.
is partially supported by the Spanish grant MINECO-16-FPA2015-63667-P,
the Ramon y Cajal fellowship RYC-2012-10370 and GRUPIN18-174 research
grant from Principado de Asturias. The work of S.M. is funded by the
Deutsche Forschungsgemeinschaft (DFG, German Research Foundation)
under Emmy Noether Programme grant no. MO 3013/1-1 and under Germany's
Excellence Strategy - EXC-2111 - 390814868. 
\end{acknowledgments}

\onecolumngrid

\appendix

\section{Geometric quantities and their perturbative expansion\label{sec:Geometric quantities}}

We write $E_{A}^{\;i}=\delta_{A}^{\;i}+H_{A}^{\;i}$ for the inverse
vielbein, and expand the relevant geometric quantities in $H$. For
the inverse metric $G^{ij}=E_{A}^{\;i}\delta^{AB}E_{B}^{\;j}$ and
volume element $\sqrt{G}=\left|E\right|=\left|\text{det}\left(E_{A}^{\;i}\right)\right|$
we find 
\begin{align}
G^{ij}= & \delta^{ij}+2H^{(ij)}+H_{A}^{\;i}H^{Aj}\label{eq:48}\\
= & \delta^{ij}+\delta G^{ij},\nonumber \\
\sqrt{G}= & 1-H_{A}^{\;A}+\frac{1}{2}H_{A}^{\;A}H_{B}^{\;B}+\frac{1}{2}H_{A}^{\;B}H_{B}^{\;A}+O\left(H^{3}\right),\nonumber \\
\log\sqrt{G}= & -H_{A}^{\;A}+\frac{1}{2}H_{A}^{\;B}H_{B}^{\;A}+O\left(H^{3}\right),\nonumber 
\end{align}
where, in expanded expressions, all index manipulations are trivial,
and in particular, there is no difference between coordinate indices
$i,j$ and $SO\left(2\right)_{L}$ indices $A,B$. Note that the strain
used in the main text is given by $u_{ij}=\left(G_{ij}-\delta_{ij}\right)/2=-H_{(ij)}+O\left(H^{2}\right)$.
We use the notation $\varepsilon^{\mu\nu\rho}$ for the totally anti-symmetric
(pseudo) tensor, normalized such that $\varepsilon^{xyt}=1/\sqrt{G}$,
as well as $\varepsilon^{ij}=\varepsilon^{ijt}$. 

The non-relativistic spin connection used in the main text is the
$SO\left(2\right)_{L}$ connection
\begin{align}
\omega_{t}= & \frac{1}{2}\varepsilon^{AB}E_{Ai}\partial_{t}E_{B}^{\;i}\label{eq:49}\\
= & -\frac{1}{2}\partial_{t}\left(\varepsilon^{AB}H_{AB}\right)-\frac{1}{2}\varepsilon^{AB}H_{iA}\partial_{t}H_{B}^{\;i}+O\left(H^{3}\right),\nonumber \\
\omega_{j}= & \frac{1}{2}\left(\varepsilon^{AB}E_{Ai}\partial_{j}E_{B}^{\;i}-\frac{1}{E}\varepsilon^{kl}\partial_{k}G_{lj}\right)\nonumber \\
= & -\frac{1}{2}\partial_{j}\left(\varepsilon^{AB}H_{AB}\right)-\partial_{\bot}^{l}H_{(lj)}-\frac{1}{2}\varepsilon^{AB}H_{iA}\partial_{j}H_{B}^{\;i}+O\left(H^{3}\right),\nonumber 
\end{align}
where $\partial_{\bot}^{l}=\varepsilon^{lk}\partial_{k}$, which is
obtained naturally within Newton-Cartan geometry \cite{moroz2015effective,gromov2016boundary}.
This connection is torsion-full, but has a vanishing ``reduced torsion''
\cite{bradlyn2015low}. In the main text, a term $B/2m$ was implicitly
added to $\omega_{t}$, but here we will add it explicitly when writing
expressions for $S_{\text{eff}}$ and $S_{\text{ind}}$. Such a term
appears in the presence of an additional background field $E_{0}^{\;i}$
which couples to momentum density $P_{i}$ \cite{bradlyn2015low,moroz2015effective},
and can be identified with $G^{ij}A_{j}/m$ in a Galilean invariant
system, where $P_{i}=mG_{ij}J^{j}$. The Ricci scalar is given by
\begin{align}
R= & 2\varepsilon^{ij}\partial_{i}\omega_{j}\\
= & 2\partial_{\bot}^{i}\partial_{\bot}^{j}H_{ij}+O\left(H^{2}\right)\nonumber \\
= & -2\left(\partial^{i}\partial^{j}-\partial^{2}\delta^{ij}\right)H_{ij}+O\left(H^{2}\right).\nonumber 
\end{align}

\section{Odd viscosity at non-zero wave-vector: generalities}

\subsection{Definition and $T$ symmetry\label{subsec: B.1}}

We define the viscosity tensor as the linear response of stress to
strain rate 
\begin{align}
 & T^{ij}\left(t,\mathbf{x}\right)=\int\text{d}t\text{d}^{2}\mathbf{x}'\eta^{ij,kl}\left(t,\mathbf{x},t',\mathbf{x}'\right)\partial_{t'}H_{kl}\left(t',\mathbf{x}'\right),
\end{align}
where 
\begin{align}
 & \eta^{[ij],kl}=0=\eta^{ij,[kl]}.\label{eq:p1}
\end{align}
In a translationally invariant system we can pass to Fourier components
$T^{ij}\left(\omega,\mathbf{q}\right)=i\omega\eta^{ij,kl}\left(\omega,\mathbf{q}\right)H_{kl}\left(\omega,\mathbf{q}\right)$.
By definition, $\eta^{ij,kl}\left(t,\mathbf{x},t',\mathbf{x}'\right)$
is real, and therefore 
\begin{align}
\eta^{ij,kl}\left(\omega,\mathbf{q}\right)= & \eta^{ij,kl}\left(-\omega,-\mathbf{q}\right)^{*}.\label{eq:p5}
\end{align}
Under time reversal $T$, 
\begin{align}
 & \eta^{ij,kl}\left(\omega,\mathbf{q}\right)\mapsto\eta_{T}^{ij,kl}\left(\omega,\mathbf{q}\right)=\eta^{kl,ij}\left(\omega,-\mathbf{q}\right).
\end{align}
The even and odd viscosities are then defined by $\eta_{\text{e},\text{o}}=\left(\eta\pm\eta_{T}\right)/2$,
and satisfy $\left(\eta_{\text{e},\text{o}}\right)_{T}=\pm\eta_{\text{e},\text{o}}$.
More explicitly, 
\begin{align}
 & \eta_{\text{e}}^{ij,kl}\left(\omega,\mathbf{q}\right)=+\eta_{\text{e}}^{kl,ij}\left(\omega,-\mathbf{q}\right),\\
 & \eta_{\text{o}}^{ij,kl}\left(\omega,\mathbf{q}\right)=-\eta_{\text{o}}^{kl,ij}\left(\omega,-\mathbf{q}\right).\nonumber 
\end{align}
We will see below that in isotropic (or $SO\left(2\right)$ invariant)
systems $\eta$ is even in $\mathbf{q}$, so that 
\begin{align}
 & \eta_{\text{e}}^{ij,kl}\left(\omega,\mathbf{q}\right)=+\eta_{\text{e}}^{kl,ij}\left(\omega,\mathbf{q}\right),\label{eq:p3}\\
 & \eta_{\text{o}}^{ij,kl}\left(\omega,\mathbf{q}\right)=-\eta_{\text{o}}^{kl,ij}\left(\omega,\mathbf{q}\right),\label{eq:p2}
\end{align}
which is identical to the definition of $\eta_{\text{e},\text{o}}$
at $\mathbf{q}=0$ \cite{PhysRevLett.75.697,Avron1998,PhysRevB.86.245309,hoyos2014hall,PhysRevE.89.043019}. 

\subsection{$SO\left(2\right)$ and $P$ symmetries}

Complex tensors satisfying \eqref{eq:p1} and \eqref{eq:p2}, in $2$
spatial dimensions, form a vector space $V\cong\mathbb{C}^{3}$ which
can be spanned by \cite{Avron1998} 
\begin{align}
 & \sigma^{ab}=2\sigma^{[a}\otimes\sigma^{b]},\;\;a,b=0,x,z,\label{eq:29-1}
\end{align}
where $\sigma^{x},\sigma^{z}$ are the symmetric Pauli matrices, and
$\sigma^{0}$ is the identity matrix. Thus every odd viscosity tensor
can be written as 
\begin{align}
\eta_{\text{o}}\left(\omega,\mathbf{q}\right) & =\eta_{xz}\left(\omega,\mathbf{q}\right)\sigma^{xz}+\eta_{x0}\left(\omega,\mathbf{q}\right)\sigma^{x0}+\eta_{z0}\left(\omega,\mathbf{q}\right)\sigma^{z0},\label{eq:30-2}
\end{align}
with complex coefficients $\eta_{ab}\left(\omega,\mathbf{q}\right)$.
Under a rotation $R=e^{i\alpha\left(i\sigma^{y}\right)}\in SO\left(2\right)$
the metric perturbation and stress tensor transform as
\begin{align}
 & H_{ij}\left(\omega,\mathbf{q}\right)\mapsto R_{\;i}^{k}R_{\;j}^{l}H_{kl}\left(\omega,R^{-1}\cdot\mathbf{q}\right),\\
 & T^{ij}\left(\omega,\mathbf{q}\right)\mapsto R_{\;k}^{i}R_{\;k}^{j}T^{kl}\left(\omega,R^{-1}\cdot\mathbf{q}\right),\nonumber 
\end{align}
where $\left(R\cdot\mathbf{q}\right)^{i}=R_{\;j}^{i}q^{j}$. The same
transformation rules apply for $R\in O\left(2\right)$, which defines
the parity transformation $P$, in flat space. It follows that 
\begin{align}
 & \eta^{ij,kl}\left(\omega,\mathbf{q}\right)\mapsto R_{\;i'}^{i}R_{\;j'}^{j}R_{\;k'}^{k}R_{\;l'}^{l}\eta^{i'j',k'l'}\left(\omega,R^{-1}\cdot\mathbf{q}\right)\label{eq:32-1}
\end{align}
under $O\left(2\right)$, which is compatible with \eqref{eq:p1},
and the decomposition $\eta=\text{\ensuremath{\eta}}_{\text{o}}+\eta_{\text{e}}$.
In particular, equation \eqref{eq:32-1} shows that the viscosity
tensor is $P$-even, or more accurately, a tensor under $P$ rather
than a pseudo-tensor. In an $SO\left(2\right)$-invariant system,
the viscosity tensor will also be $SO\left(2\right)$-invariant
\begin{align}
 & \eta^{ij,kl}\left(\omega,\mathbf{q}\right)=R_{\;i'}^{i}R_{\;j'}^{j}R_{\;k'}^{k}R_{\;l'}^{l}\eta^{i'j',k'l'}\left(\omega,R^{-1}\cdot\mathbf{q}\right),\;R\in SO\left(2\right).\label{eq:33-1}
\end{align}
Note that this holds even when $SO\left(2\right)$ symmetry is \textit{spontaneously}
broken, as in $\ell$-wave SFs. At $\mathbf{q}=\mathbf{0}$, there
is a unique tensor satisfying \eqref{eq:33-1}, namely 
\begin{align}
\left(\sigma^{xz}\right)^{ij,kl} & =-\frac{1}{2}\left(\varepsilon^{ik}\delta^{jl}+\varepsilon^{jk}\delta^{il}+\varepsilon^{il}\delta^{jk}+\varepsilon^{jl}\delta^{ik}\right),\label{34}
\end{align}
leaving a single odd viscosity coefficient $\eta_{xz}\left(\omega\right)=\eta_{\text{o}}^{\left(1\right)}\left(\omega\right)$
\cite{PhysRevLett.75.697,Avron1998,PhysRevB.86.245309,hoyos2014hall,PhysRevE.89.043019}.

A non-zero $\mathbf{q}$, however, along with the  tensors $\delta^{ij}$
and $\varepsilon^{ij}$, can be used to construct additional $SO\left(2\right)$-invariant
odd viscosity tensors, beyond $\sigma^{xz}$. From the data $\mathbf{q},\delta^{ij},\varepsilon^{ij}$,
three linearly independent, symmetric, rank-2 tensors can be constructed,
which we take to be 
\begin{align}
\left(\tau^{0}\right)^{ij}= & q^{2}\delta^{ij},\label{eq:35-1}\\
\left(\tau^{x}\right)^{ij}= & -2q_{\bot}^{(i}q^{j)}/q^{2},\nonumber \\
\left(\tau^{z}\right)^{ij}= & 2q^{i}q^{j}/q^{2}-\delta^{ij},\nonumber 
\end{align}
where $q_{\bot}^{i}=\varepsilon^{ij}q_{j}$. The notation above is
due to the relation
\begin{align}
\begin{pmatrix}\tau^{x}\\
\tau^{z}
\end{pmatrix} & =\begin{pmatrix}\cos2\theta & -\sin2\theta\\
\sin2\theta & \cos2\theta
\end{pmatrix}\begin{pmatrix}\sigma^{x}\\
\sigma^{z}
\end{pmatrix}\label{eq:35}\\
 & =\frac{1}{q^{2}}\begin{pmatrix}q_{x}^{2}-q_{y}^{2} & -2q_{x}q_{y}\\
2q_{x}q_{y} & q_{x}^{2}-q_{y}^{2}
\end{pmatrix}\begin{pmatrix}\sigma^{x}\\
\sigma^{z}
\end{pmatrix},\nonumber 
\end{align}
where $\theta=\text{arg}\left(\mathbf{q}\right)$, so that $\tau^{x},\tau^{z}$
are a rotated version of $\sigma^{x},\sigma^{z}$. Moreover, all three
$\tau$s are $SO\left(2\right)$-invariant, $\tau^{ij}\left(\mathbf{q}\right)=R_{\;i'}^{i}R_{\;j'}^{j}\tau^{i'j'}\left(R^{-1}\cdot\mathbf{q}\right)$,
and can therefore be used to construct three $SO\left(2\right)$-invariant
odd viscosity tensors 
\begin{align}
 & \tau^{ab}=2\tau^{[a}\otimes\tau^{b]},\;a,b=0,x,z,
\end{align}
which form a basis for $V$. Any odd viscosity tensor (at $\mathbf{q}\neq\mathbf{0}$)
can then be written as 
\begin{align}
\eta_{\text{o}}\left(\omega,\mathbf{q}\right) & =\eta_{\text{o}}^{\left(1\right)}\left(\omega,\mathbf{q}\right)\tau^{xz}+\eta_{\text{o}}^{\left(2\right)}\left(\omega,\mathbf{q}\right)\tau^{0x}+\eta_{\text{o}}^{\left(3\right)}\left(\omega,\mathbf{q}\right)\tau^{0z}.
\end{align}
Furthermore, for an $SO\left(2\right)$-invariant $\eta_{\text{o}}$,
the coefficients $\eta_{\text{o}}^{\left(1\right)},\eta_{\text{o}}^{\left(2\right)},\eta_{\text{o}}^{\left(3\right)}$
depend on $\mathbf{q}$ through its norm, owing to the $SO\left(2\right)$-invariance
of $\tau^{ab}$. We therefore arrive at the general form of an $SO\left(2\right)$-invariant
odd viscosity tensor, 
\begin{align}
\eta_{\text{o}}\left(\omega,\mathbf{q}\right)= & \eta_{\text{o}}^{\left(1\right)}\left(\omega,q^{2}\right)\tau^{xz}+\eta_{\text{o}}^{\left(2\right)}\left(\omega,q^{2}\right)\tau^{0x}+\eta_{\text{o}}^{\left(3\right)}\left(\omega,q^{2}\right)\tau^{0z}.\label{eq:39-2}
\end{align}
In particular, we see that $\eta_{\text{o}}$ is even in $\mathbf{q}$
(and the same applies also to the even viscosity $\eta_{\text{e}}$).
To determine the small $\omega,\mathbf{q}$ behavior of the coefficients
we change to the $\mathbf{q}$-independent basis of $\sigma$s, 
\begin{align}
\eta_{\text{o}}\left(\omega,\mathbf{q}\right)= & \eta_{\text{o}}^{\left(1\right)}\left(\omega,q^{2}\right)\sigma^{xz}\label{eq:39-1}\\
 & +\left[\eta_{\text{o}}^{\left(2\right)}\left(\omega,q^{2}\right)\left(q_{x}^{2}-q_{y}^{2}\right)+\eta_{\text{o}}^{\left(3\right)}\left(\omega,q^{2}\right)\left(2q_{x}q_{y}\right)\right]\sigma^{0x}\nonumber \\
 & +\left[\eta_{\text{o}}^{\left(2\right)}\left(\omega,q^{2}\right)\left(-2q_{x}q_{y}\right)+\eta_{\text{o}}^{\left(3\right)}\left(\omega,q^{2}\right)\left(q_{x}^{2}-q_{y}^{2}\right)\right]\sigma^{0z}.\nonumber 
\end{align}
In gapped systems (such as QH states) $\eta_{\text{o}}$ will be regular
around $\omega=0=q$, and so will the coefficients $\eta_{\text{o}}^{\left(1\right)},\eta_{\text{o}}^{\left(2\right)},\eta_{\text{o}}^{\left(3\right)}$.
In gapless systems (such as $\ell$-wave SFs) there will be a singularity
at $\omega=0=q$, but the limit $q\rightarrow0$ at $\omega\neq0$
will be regular. In both cases, the limit $q\rightarrow0$ at $\omega\neq0$
of \eqref{eq:39-1} reduces to the known result $\eta_{\text{o}}\left(\omega,\mathbf{0}\right)=\eta_{\text{o}}^{\left(1\right)}\left(\omega,0\right)\sigma^{xz}$
\cite{PhysRevLett.75.697,Avron1998,PhysRevB.86.245309,hoyos2014hall,PhysRevE.89.043019}. 

\subsection{$PT$ symmetry}

The combination $PT$ of parity and time-reversal is a symmetry in
any system in which $T$ is broken (perhaps spontaneously) due to
some kind of angular momentum, as in QH states, $\ell$-wave SFs,
and active chiral fluids \cite{banerjee2017odd}. Here we consider
the implications of $PT$ symmetry on \eqref{eq:39-1}. 

From the definition \eqref{eq:35-1} it is clear that $\tau^{0}$
and $\tau^{z}$ are $P$-even, while $\tau^{x}$ is $P$-odd. Therefore,
$\tau^{xz},\tau^{0x}$ are $P$-odd while $\tau^{0z}$ is $P$-even
(and all three are $T$-even). Since $\eta_{\text{o}}$ is $T$-odd
and $P$-even, and using \eqref{eq:39-2}, it follows that $\eta_{\text{o}}^{\left(1\right)}$
and $\eta_{\text{o}}^{\left(2\right)}$ are $P,T$-odd, while $\eta_{\text{o}}^{\left(3\right)}$
is $T$-odd but $P$-even. In particular, $\eta_{\text{o}}^{\left(3\right)}$
is $PT$-odd, and must vanish in $PT$-symmetric systems. The odd
viscosity tensor in $SO\left(2\right)$ and $PT$ symmetric systems
is therefore given by 
\begin{align}
\eta_{\text{o}}\left(\omega,\mathbf{q}\right)= & \eta_{\text{o}}^{\left(1\right)}\left(\omega,q^{2}\right)\sigma^{xz}\label{eq:39-1-1}\\
 & +\eta_{\text{o}}^{\left(2\right)}\left(\omega,q^{2}\right)\left[\left(q_{x}^{2}-q_{y}^{2}\right)\sigma^{0x}-2q_{x}q_{y}\sigma^{0z}\right].\nonumber 
\end{align}
This form is confirmed by previous results for QH states \cite{abanov2014electromagnetic},
and by the results presented in Sec.\ref{sec: induced action and linear response} for CSFs.
The same form is obtained at $\mathbf{q}=\mathbf{0}$, but in the
presence of vector, or pseudo-vector, anisotropy $\mathbf{b}$, in
which case we find 
\begin{align}
\eta_{\text{o}}\left(\omega\right)= & \eta_{\text{o}}^{\left(1\right)}\left(\omega\right)\sigma^{xz}\label{eq:39-1-1-1}\\
 & +\eta_{\text{o}}^{\left(2\right)}\left(\omega\right)\left[\left(b_{x}^{2}-b_{y}^{2}\right)\sigma^{0x}-2b_{x}b_{y}\sigma^{0z}\right],\nonumber 
\end{align}
which explains the tensor structure found in \cite{PhysRevB.99.045141,PhysRevB.99.035427}.

\subsection{Frequency dependence and reality conditions\label{subsec: B.4}}

In closed and clean systems, like the $\ell$-wave SFs discussed
in this paper, the viscosity can be obtained from an induced action
\begin{align}
S_{\text{ind}} & \supset\frac{1}{2}\int\text{d}t\text{d}t'\text{d}^{2}\mathbf{x}\text{d}^{2}\mathbf{x}'H_{ij}\left(t,\mathbf{x}\right)\eta^{ij,kl}\left(t-t',\mathbf{x}-\mathbf{x}'\right)\partial_{t'}H_{kl}\left(t',\mathbf{x}'\right)\label{eq:38}\\
 & =\frac{1}{2}\int\frac{\text{d}\omega}{2\pi}\frac{\text{d}^{2}\mathbf{q}}{\left(2\pi\right)^{2}}H_{ij}\left(-\omega,-\mathbf{q}\right)i\omega\eta^{ij,kl}\left(\omega,\mathbf{q}\right)H_{kl}\left(\omega,\mathbf{q}\right).\nonumber 
\end{align}
As a result, $\eta$ satisfies the additional property, 
\begin{align}
\eta^{ij,kl}\left(\omega,\mathbf{q}\right)= & -\eta^{kl,ij}\left(-\omega,-\mathbf{q}\right),\label{eq:p4}
\end{align}
which, along with \eqref{eq:p2}-\eqref{eq:p3} and the fact that
$\eta$ is even in $\mathbf{q}$, implies that $\eta_{\text{o}}$
($\eta_{\text{e}}$) is even (odd) in $\omega$,
\begin{align}
\eta_{\text{e}}^{ij,kl}\left(\omega,\mathbf{q}\right)= & -\eta_{\text{e}}^{ij,kl}\left(-\omega,\mathbf{q}\right),\label{eq:44}\\
\eta_{\text{o}}^{ij,kl}\left(\omega,\mathbf{q}\right)= & +\eta_{\text{o}}^{ij,kl}\left(-\omega,\mathbf{q}\right).\nonumber 
\end{align}
 This result, along with \eqref{eq:p5} and the fact that $\eta$
is even in $\mathbf{q}$, implies that $\eta_{\text{o}}$ ($\eta_{\text{e}}$)
is real (imaginary),
\begin{align}
\eta_{\text{e}}^{ij,kl}\left(\omega,\mathbf{q}\right)\in & i\mathbb{R},\label{eq:45}\\
\eta_{\text{o}}^{ij,kl}\left(\omega,\mathbf{q}\right)\in & \mathbb{R}.\nonumber 
\end{align}
These general properties are satisfied by the odd viscosity tensor
 computed in this paper. These are also compatible with the examples
worked out in \cite{PhysRevB.86.245309}, as well with viscosity-conductivity
relations that hold in Galilean invariant systems (in conjugation
with known properties of the conductivity) \cite{hoyos2012hall,PhysRevB.86.245309,hoyos2014effective}. 

We note that some care is required when interpreting \eqref{eq:44}-\eqref{eq:45}
around singularities of $\eta$. For example, the first equation in
\eqref{eq:44} \textit{naively} implies that $\eta_{\text{e}}\left(0,\mathbf{q}\right)=0$,
which in particular implies that the bulk and shear viscosities $\eta_{\text{e}}\left(0,\mathbf{0}\right)=\zeta\sigma^{0}\otimes\sigma^{0}+\eta^{\text{s}}\left(\sigma^{x}\otimes\sigma^{z}+\sigma^{z}\otimes\sigma^{x}\right)$
 vanish in the closed, clean, case. This however, is not quite correct,
due to a possible singularity of $\eta_{\text{e}}$ at $\omega=0$,
as well as the usual infinitesimal imaginary part of $\omega$ required
to obtain the retarded response. For example, for free fermions, reference
\cite{PhysRevB.86.245309} finds $\eta^{\text{s}}\left(\omega,\mathbf{0}\right)\sim\frac{i}{\omega+i\epsilon}=\pi\delta\left(\omega\right)+i\text{PV}\frac{1}{\omega}$
(where $\text{PV}$ is the principle value), which has an infinite
real part at $\omega=0$, in analogy with the Drude behavior of the
conductivity.  

\subsection{Odd viscosity from Gaussian integration: a technical result\label{subsec:Odd-viscosity-from}}

We now restrict attention to CSFs. The effective Lagrangian,
perturbatively expanded to second order, and in the absence of the
$U\left(1\right)$ background, takes the form 
\begin{align}
\mathcal{L}_{\text{eff}}= & \frac{1}{2}\theta\mathcal{G}^{-1}\theta+\mathcal{V}\theta+\mathcal{C},\label{eq:r1}
\end{align}
where the Green's function $\mathcal{G}$ is independent of $H$,
the vertex $\mathcal{V}$ is linear in $H$, and the contact term
$\mathcal{C}$ is quadratic in $H$. Performing Gaussian integration
over $\theta$ yields the induced Lagrangian 
\begin{align}
\text{\ensuremath{\mathcal{L}}}_{\text{ind}}= & -\frac{1}{2}\mathcal{V}\mathcal{G}\mathcal{V}+\mathcal{C},\label{eq:44-1}
\end{align}
and comparing with \eqref{eq:38} one can read off $\eta_{\text{o}}$.
In Appendix \ref{subsec:Effective-action-and} we write explicit expressions
for a Galilean invariant $\mathcal{L}_{\text{eff}}$, which we then
expand to obtain explicit expressions for $\mathcal{G}^{-1},\mathcal{V},\mathcal{C}$.
Appendix \ref{subsec:Obtaining--from} then describes the resulting
$\mathcal{L}_{\text{ind}}$. Here we take a complementary approach
and obtain the general form of $\eta_{\text{o}}$ from \eqref{eq:44-1},
using the formalism developed above, based only on $SO\left(2\right)$
and $PT$ symmetries. 

The motivation for the analysis in this appendix is the following.
The power counting \eqref{eq: power count}  is designed such that
the $O\left(p^{n}\right)$ Lagrangian $\mathcal{L}_{n}\subset\mathcal{L}_{\text{eff}}$
produces $O\left(p^{n}\right)$ contributions to $\mathcal{L}_{\text{ind}}$.
Therefore, naively, one expects the $O\left(q^{2}\right)$ odd viscosity
to depend on $\mathcal{L}_{0},\mathcal{L}_{2},$ and $\mathcal{L}_{3}$
(since $\mathcal{L}_{1}=0$). Using the notation $\eta_{\text{o}}=\eta_{\mathcal{V}}+\eta_{\mathcal{C}}$
for the parts of $\eta_{\text{o}}$ due to $-\mathcal{V}\mathcal{G}\mathcal{V}/2$
and $\mathcal{C}$, respectively, the result of this appendix is that
$\eta_{\mathcal{V}}$, to $O\left(q^{2}\right)$, is actually independent
of $\mathcal{L}_{3}$. 

We now describe the details. For $\eta_{\mathcal{C}}$, we cannot
do better than the general discussion thus far - it is given by \eqref{eq:39-1},
with $\eta_{\text{o}}^{\left(3\right)}=0$, and both $\eta_{\text{o}}^{\left(1\right)},\eta_{\text{o}}^{\left(2\right)}$
are real and regular at $\omega=0=q$, since $\mathcal{L}_{\text{eff}}$,
and $\mathcal{C}$ in particular, are obtained by integrating out
gapped degrees of freedom (the Higgs modes and the fermion $\psi$).
For $\eta_{\mathcal{V}}$, however, we can do better. We first write
more explicitly 
\begin{align}
\theta\mathcal{G}^{-1}\theta & =\frac{1}{2}\theta\left(-\omega,-\mathbf{q}\right)\mathcal{G}^{-1}\left(\omega,\mathbf{q}\right)\theta\left(\omega,\mathbf{q}\right),\label{eq:r2}\\
\mathcal{V}\theta & =\theta\left(-\omega,-\mathbf{q}\right)V^{ij}\left(\omega,\mathbf{q}\right)H_{ij}\left(\omega,\mathbf{q}\right).\nonumber 
\end{align}
Based on $SO\left(2\right)$ and $PT$ symmetries, the objects $\mathcal{G}^{-1},V^{ij}$
take the forms 
\begin{align}
\mathcal{G}^{-1}\left(\omega,\mathbf{q}\right)= & D\left(\omega^{2},q^{2}\right),\label{eq:r3}\\
V^{ij}\left(\omega,\mathbf{q}\right)= & i\omega a\left(\omega^{2},q^{2}\right)\left(\rho^{0}\right)^{ij}+i\omega b\left(\omega^{2},q^{2}\right)\left(\rho^{z}\right)^{ij}+s_{\theta}c\left(\omega^{2},q^{2}\right)\left(\rho^{x}\right)^{ij},\nonumber 
\end{align}
where 
\begin{align}
\left(\rho^{0}\right)^{ij}= & \delta^{ij},\label{eq:35-1-1-1-1}\\
\left(\rho^{x}\right)^{ij}= & q_{\bot}^{(i}q^{j)},\nonumber \\
\left(\rho^{z}\right)^{ij}= & q^{i}q^{j},\nonumber 
\end{align}
are, in this context, more convenient than the $\tau$s \eqref{eq:35-1},
and $a,b,c,D$ are general functions of their arguments which are
$P,T$-even, real, and regular at $\omega=0=q$, as follows from the
same properties of $\mathcal{L}_{\text{eff}}$. In particular, we
will use the following expansions 
\begin{align}
a\left(0,q^{2}\right) & =a_{0}+a_{1}q^{2}+O\left(q^{4}\right),\label{eq:r4}\\
b\left(0,q^{2}\right) & =b_{0}+O\left(q^{2}\right),\nonumber \\
c\left(0,q^{2}\right) & =c_{0}+c_{1}q^{2}+O\left(q^{4}\right),\nonumber \\
D\left(0,q^{2}\right) & =D_{1}q^{2}+D_{2}q^{4}+O\left(q^{6}\right),\nonumber 
\end{align}
where $D_{0}=0$ because $\theta$ enters $\mathcal{L}_{\text{eff}}$
only through its derivatives. The odd viscosity $\eta_{\mathcal{V}}$
is then given by 
\begin{align}
\eta_{\mathcal{V}}\left(\omega,\mathbf{q}\right)= & -\frac{1}{2i\omega}\frac{V\left(-\omega,-\mathbf{q}\right)\otimes V\left(\omega,\mathbf{q}\right)-V\left(\omega,\mathbf{q}\right)\otimes V\left(-\omega,-\mathbf{q}\right)}{D\left(\omega,\mathbf{q}\right)}\\
= & \frac{2s_{\theta}c\left(\omega^{2},q^{2}\right)}{D\left(\omega^{2},q^{2}\right)}\left[a\left(\omega^{2},q^{2}\right)\rho^{0x}+b\left(\omega^{2},q^{2}\right)\rho^{zx}\right],\nonumber 
\end{align}
which is of the form \eqref{eq:39-1}, with $\eta_{\text{o}}^{\left(3\right)}=0$
and 
\begin{align}
\eta_{\mathcal{V}}^{\left(1\right)}\left(\omega,q^{2}\right)= & -\frac{s_{\theta}c\left(\omega^{2},q^{2}\right)}{2D\left(\omega^{2},q^{2}\right)}b\left(\omega^{2},q^{2}\right)q^{4},\\
\eta_{\text{\ensuremath{\mathcal{V}}}}^{\left(2\right)}\left(\omega,q^{2}\right)= & \frac{s_{\theta}c\left(\omega^{2},q^{2}\right)}{D\left(\omega^{2},q^{2}\right)}\left[a\left(\omega^{2},q^{2}\right)+b\left(\omega^{2},q^{2}\right)q^{2}\right].\nonumber 
\end{align}
Setting $\omega=0$ and expanding in $q$, we find
\begin{align}
\eta_{\text{\ensuremath{\mathcal{V}}}}^{\left(1\right)}\left(0,q^{2}\right)= & -\frac{s_{\theta}c_{0}b_{0}}{2D_{1}}q^{2}+O\left(q^{4}\right),\label{eq:22-2}\\
\eta_{\text{\ensuremath{\mathcal{V}}}}^{\left(2\right)}\left(0,q^{2}\right)= & \frac{s_{\theta}}{D_{1}}\left[a_{0}c_{0}q^{-2}+\left(a_{0}c_{1}+a_{1}c_{0}+b_{0}c_{0}-a_{0}c_{0}\frac{D_{2}}{D_{1}}\right)\right]+O\left(q^{2}\right).\nonumber 
\end{align}
Having identified the coefficients $a_{0},a_{1},b_{0},c_{0},c_{1},D_{1},D_{2}$
that determine $\eta_{\mathcal{V}}$ to $O\left(q^{2}\right)$, we
now determine the order in the derivative expansion  of $\mathcal{L}_{\text{eff}}$
in which these enter. Explicitly, the above coefficients are defined
by 
\begin{align}
\mathcal{L}_{\text{eff}}\supset & \frac{1}{2}\theta\left(-\omega,-\mathbf{q}\right)\left(D_{1}q^{2}+D_{2}q^{4}\right)\theta\left(\omega,\mathbf{q}\right)\label{eq:130}\\
 & +\theta\left(-\omega,-\mathbf{q}\right)\left[i\omega\left(a_{0}+a_{1}q^{2}\right)\delta^{ij}+i\omega b_{0}q^{i}q^{j}+s_{\theta}\left(c_{0}+c_{1}q^{2}\right)q^{i}q_{\bot}^{j}\right]H_{ij}\left(\omega,\mathbf{q}\right).\nonumber 
\end{align}
We see that $c_{1}$ enters $\mathcal{L}_{\text{eff}}$ at $O\left(p^{3}\right)$,
while all other coefficients enter at a lower order, and come from
$\mathcal{L}_{0},\mathcal{L}_{2}$. In particular, $\eta_{\text{\ensuremath{\mathcal{V}}}}^{\left(1\right)}$
in \eqref{eq:22-2} is independent of $\mathcal{L}_{3}$. Even though
$c_{1}$ is the coefficient of an $O\left(p^{3}\right)$ term, it
is actually due to $\mathcal{L}_{2}$. Using \eqref{eq:49} we identify
$c_{0}\theta q^{2}q^{i}q_{\bot}^{j}H_{ij}=-s_{\theta}c_{1}\partial^{i}\theta\partial^{2}\omega_{i}$,
which must be a part of 
\begin{align}
 & \frac{c_{1}}{2}\left(\partial_{i}\theta-A_{i}-s_{\theta}\omega_{i}\right)\partial^{2}\left(\partial_{i}\theta-A_{i}-s_{\theta}\omega_{i}\right).
\end{align}
This is an $O\left(p^{2}\right)$ term, and in fact comes from $\mathcal{L}_{2}^{\left(2\right)}\subset\mathcal{L}_{2}$,
see \eqref{eq:54}. Thus, both $\eta_{\text{\ensuremath{\mathcal{V}}}}^{\left(1\right)},\eta_{\text{\ensuremath{\mathcal{V}}}}^{\left(2\right)}$
in \eqref{eq:22-2} are completely independent of $\mathcal{L}_{3}$.

\section{Effective action and its perturbative expansion\label{subsec:Effective-action-and}
}

\subsection{Zeroth order}

It is useful to write the zeroth order scalar $X$ as 
\begin{align}
X= & \left(\partial_{t}\theta-\mathcal{A}_{t}-\frac{s_{\theta}}{2m}B\right)-\frac{1}{2m}G^{ij}\left(\partial_{i}\theta-\mathcal{A}_{i}\right)\left(\partial_{j}\theta-\mathcal{A}_{j}\right),\label{eq:69}
\end{align}
where
\begin{align}
\mathcal{A}_{\mu} & =A_{\mu}+s_{\theta}\omega_{\mu}.\label{eq:70}
\end{align}
We will also use $\mathcal{B}=B+\frac{s_{\theta}}{2}R,\;\mathcal{E}_{i}=E_{i}+s_{\theta}E_{\omega,i}$
for the magnetic and electric fields obtained from $\mathcal{A}_{\mu}$,
where $E_{\omega,i}=\partial_{t}\omega_{i}-\partial_{i}\omega_{t}$.
Expanding $\mathcal{L}_{0}=P\left(X\right)$  to second order in
the fields, one finds (up to total derivatives)
\begin{align}
\sqrt{G}\mathcal{L}_{0}= & \frac{1}{2}\frac{n_{0}}{m}\theta\left[\partial^{2}-c_{s}^{-2}\partial_{t}^{2}\right]\theta\label{eq:51}\\
 & +\left[-\frac{n_{0}}{m}\left(\partial_{i}\mathcal{A}^{i}-c_{s}^{-2}\partial_{t}\left(\mathcal{A}_{t}+\frac{s_{\theta}}{2m}B\right)\right)-n_{0}\partial_{t}\sqrt{G}\right]\theta\nonumber \\
 & +\left[-n_{0}\sqrt{G}\mathcal{A}_{t}-\frac{1}{2}\frac{n_{0}}{m}\left(\mathcal{A}^{2}-c_{s}^{-2}\left(\mathcal{A}_{t}+\frac{s_{\theta}}{2m}B\right)^{2}\right)+P_{0}\sqrt{G}\right]\nonumber \\
= & \frac{1}{2}\theta\mathcal{G}^{-1}\theta+\mathcal{V}\theta+\mathcal{C},\nonumber 
\end{align}
where $\partial^{2}=\partial^{i}\partial_{i},\;\mathcal{A}^{2}=\mathcal{A}_{i}\mathcal{A}^{i}$,
and we defined the inverse Green's function $\mathcal{G}^{-1}$, vertex
$\mathcal{V}$, and contact terms $\mathcal{C}$, respectively. These
are used in Appendix \ref{subsec:Obtaining--from} below to obtain
$S_{\text{ind}}$. 

In \eqref{eq:51}, the geometric objects $\sqrt{G}$ and $\omega_{\mu}$
should be interpreted as expanded to the required order according
to \eqref{eq:48}-\eqref{eq:49}. In particular, the term $-n_{0}\sqrt{G}\mathcal{A}_{t}$
includes $-s_{\theta}n_{0}\sqrt{G}\omega_{t}$, which produces the
leading contribution to $\eta_{\text{o}}^{\left(1\right)}$. To see
this, we expand 

\begin{align}
\sqrt{G}\omega_{t}= & -\frac{1}{2}\partial_{t}\left(\varepsilon^{AB}H_{AB}\right)+\frac{1}{2}\partial_{t}\left(\varepsilon^{AB}H_{AB}\right)H_{i}^{\;i}-\frac{1}{2}\varepsilon^{AB}H_{iA}\partial_{t}H_{B}^{\;i}+O\left(H^{3}\right)\label{eq:50}\\
= & -\frac{1}{2}\partial_{t}\left(\varepsilon^{AB}H_{AB}\right)-\frac{1}{2}\varepsilon^{AB}H_{Ai}\partial_{t}H_{B}^{\;i}+O\left(H^{3}\right),\nonumber 
\end{align}
which is identical to the expansion \eqref{eq:48} of $\omega_{t}$,
apart from $H_{iA}\leftrightarrow H_{Ai}$. Ignoring total derivatives,
this reduces to 
\begin{align}
\sqrt{G}\mathcal{L}_{0} & \supset-s_{\theta}n_{0}\sqrt{G}\omega_{t}\label{eq:43}\\
 & =-\frac{1}{2}s_{\theta}n_{0}\left[\partial_{t}\left(\varepsilon^{AB}H_{AB}\right)H_{i}^{\;i}-\varepsilon^{AB}\delta^{ij}H_{(Ai)}\partial_{t}H_{(Bj)}\right]+O\left(H^{3}\right)\nonumber \\
 & =\frac{1}{2}s_{\theta}n_{0}\varepsilon^{AB}H_{Ai}\partial_{t}H_{B}^{\;i}+O\left(H^{3}\right).\nonumber 
\end{align}
Comparing with \eqref{34} and \eqref{eq:38}, the second term in
the second line corresponds to $\eta_{\text{o}}^{\left(1\right)}=-s_{\theta}n_{0}/2$.
The first term in the second line depends on the anti-symmetric part
of $H$, and shows that the full expression \eqref{eq:43} actually
corresponds to a \textit{torsional} Hall (or odd) viscosity \cite{hughes2011torsional,hughes2013torsional}
$\zeta_{H}=-s_{\theta}n_{0}$, which can be read off from the third
line. The appearance of the torsional Hall viscosity at the level
of $S_{\text{eff}}$ (but not at the level of $S_{\text{ind}}$, see
Appendix \ref{subsec:Obtaining--from}) can be understood from the
mapping of \cite{PhysRevB.98.064503} of the $p$-wave SF to a Majorana
spinor in Riemann-Cartan space-time. 

\subsection{Second order\label{subsec: Second order effective action}}

The full expression for $\mathcal{L}_{2}$ is given by $\mathcal{L}_{2}=\sum_{i=1}^{6}\mathcal{L}_{2}^{\left(i\right)}$
, where \cite{son2006general}
\begin{align}
\mathcal{L}_{2}^{\left(1\right)}= & F_{1}\left(X\right)R,\label{eq:49-2}\\
\mathcal{L}_{2}^{\left(2\right)}= & F_{2}\left(X\right)\left(mK_{\;i}^{i}-\nabla^{2}\theta\right)^{2},\nonumber \\
\mathcal{L}_{2}^{\left(3\right)}= & F_{3}\left(X\right)\left\{ -m^{2}\left(G^{ij}\partial_{t}K_{ij}-K^{ij}K_{ij}\right)-m\nabla_{i}E^{i}+\frac{1}{4}F^{ij}F_{ij}\right.\nonumber \\
 & \left.+2m\left[\partial_{i}K_{\;j}^{j}-\nabla^{j}\left(K_{ji}+\frac{1}{2m}F_{ji}\right)\right]\nabla^{i}\theta+R_{ij}\nabla_{i}\theta\nabla_{j}\theta\right\} ,\nonumber \\
\mathcal{L}_{2}^{\left(4\right)}= & F_{4}\left(X\right)G^{ij}\partial_{i}X\partial_{j}X,\nonumber \\
\mathcal{L}_{2}^{\left(5\right)}= & F_{5}\left(X\right)\left[\left(\partial_{t}-\frac{1}{m}\nabla^{i}\theta\partial_{i}\right)X\right]^{2},\nonumber \\
\mathcal{L}_{2}^{\left(6\right)}= & F_{6}\left(X\right)\left(mK_{\;i}^{i}-\nabla^{2}\theta\right)\left[\left(\partial_{t}-\frac{1}{m}\nabla^{i}\theta\partial_{i}\right)X\right].\nonumber 
\end{align}
 The terms $\mathcal{L}_{2}^{\left(5\right)}$ and $\mathcal{L}_{2}^{\left(6\right)}$
were not written explicitly in \cite{son2006general} because, on
shell (on the equation of motion for $\theta$), they are proportional
to $\mathcal{L}_{2}^{\left(4\right)}$ up to $O\left(p^{4}\right)$
corrections, and can therefore be eliminated by a redefinition of
$F_{4}$. However, for the purpose of comparing the general $S_{\text{eff}}$
with the microscopic expression \eqref{eq:39}, it is convenient to
work off shell and keep all terms explicit. 

Specializing to 2+1 dimensions and expanding to second order in fields,
one finds
\begin{align}
\sqrt{G}\mathcal{L}_{2}^{\left(1\right)}= & F_{1}'\left(\mu\right)R\left(\partial_{t}\theta-\mathcal{A}_{t}-\frac{s_{\theta}}{2m}B\right).\label{eq:54}\\
\sqrt{G}\mathcal{L}_{2}^{\left(2\right)}= & F_{2}\left(\mu\right)\left[-m^{2}H_{i}^{\;i}\partial_{t}^{2}H_{j}^{\;j}+2m\partial_{t}H_{k}^{\;k}\partial^{j}\left(\partial_{j}\theta-\mathcal{A}_{j}\right)-\left(\partial_{i}\theta-\mathcal{A}_{i}\right)\partial^{i}\partial^{j}\left(\partial_{i}\theta-\mathcal{A}_{i}\right)\right],\nonumber \\
\sqrt{G}\mathcal{L}_{2}^{\left(3\right)}= & F_{3}\left(\mu\right)\left(m^{2}H^{(ij)}\partial_{t}^{2}H_{\left(ij\right)}+\frac{1}{2}B^{2}-2m\varepsilon^{ij}\omega_{i}\partial_{t}\left(\partial_{j}\theta-\mathcal{A}_{j}\right)-B\mathcal{B}\right)\nonumber \\
 & +F_{3}'\left(\mu\right)\left(\partial_{t}\theta-\mathcal{A}_{t}-\frac{s_{\theta}}{2m}B\right)\left(m^{2}\partial_{t}^{2}H_{i}^{\;i}-m\partial_{i}E^{i}\right),\nonumber \\
\sqrt{G}\mathcal{L}_{2}^{\left(4\right)}= & -F_{4}\left(\mu\right)\left(\partial_{t}\theta-\mathcal{A}_{t}-\frac{s_{\theta}}{2m}B\right)\partial^{2}\left(\partial_{t}\theta-\mathcal{A}_{t}-\frac{s_{\theta}}{2m}B\right),\nonumber \\
\sqrt{G}\mathcal{L}_{2}^{\left(5\right)}= & -F_{5}\left(\mu\right)\left(\partial_{t}\theta-\mathcal{A}_{t}-\frac{s_{\theta}}{2m}B\right)\partial_{t}^{2}\left(\partial_{t}\theta-\mathcal{A}_{t}-\frac{s_{\theta}}{2m}B\right),\nonumber \\
\sqrt{G}\mathcal{L}_{2}^{\left(6\right)}= & -F_{6}\left(\mu\right)\left[m\partial_{t}H_{i}^{\;i}+\partial^{j}\left(\partial_{j}\theta-\mathcal{A}_{j}\right)\right]\partial_{t}\left(\partial_{t}\theta-\mathcal{A}_{t}-\frac{s_{\theta}}{2m}B\right),\nonumber 
\end{align}
from which one can easily extract the second order corrections to
$\mathcal{G}^{-1},\mathcal{V},\mathcal{C}$, of \eqref{eq:51}. Note
that $\mathcal{L}_{2}^{\left(3\right)}$ includes a term $\propto\varepsilon^{ij}\omega_{i}\partial_{t}\mathcal{A}_{j}=\varepsilon^{ij}\omega_{i}\partial_{t}\left(A_{j}+s_{\theta}\omega_{j}\right)$.
Comparing with \eqref{eq:54-2} below, it is clear that distinguishing
$\mathcal{L}_{2}^{\left(3\right)}$ from $\mathcal{L}_{\text{gCS}}$
is non-trivial. This is in fact the same problem of extracting the
central charge from the Hall viscosity addressed in the main text,
but at the level of $S_{\text{eff}}$ (where $\theta$ is viewed as
a background field) rather than $S_{\text{ind}}$ (where $\theta$
has been integrated out). Accordingly, the central charge can be computed
by applying Eq.\eqref{eq:16-2} to the response functions
obtained from $S_{\text{eff}}$. Additionally, relying on LGS, one
can extract $F_{3}$ as the coefficient of $H^{(ij)}\partial_{t}^{2}H_{\left(ij\right)}$.
Both approaches produce the same central charge \eqref{eq:12-1} in
the perturbative computation of Appendix \ref{subsec:Perturbative-expansion}.

\subsection{Gravitational Chern-Simons term}

The gCS Lagrangian  is given explicitly by
\begin{align}
\mathcal{L}_{\text{gCS}} & =-\frac{c}{48\pi}\left[\left(\omega_{t}+\frac{B}{2m}\right)R-\varepsilon^{ij}\omega_{i}\partial_{t}\omega_{j}\right]\label{eq:54-2}\\
 & =-\frac{c}{48\pi}\left[\omega\text{d}\omega+\frac{1}{2m}BR\right].\nonumber 
\end{align}
Its expansion to second order in fields, using \eqref{eq:48}-\eqref{eq:49},
is
\begin{align}
\sqrt{G}\mathcal{L}_{\text{gCS}} & =-\frac{c}{48\pi}\left[\varepsilon^{AB}H_{\left(Ai\right)}\partial_{\bot}^{i}\partial_{\bot}^{j}\partial_{t}H_{\left(Bj\right)}-\frac{1}{m}A_{i}\partial_{\bot}^{i}\partial_{\bot}^{j}\partial_{\bot}^{k}H_{(jk)}\right].
\end{align}
As opposed to $\sqrt{G}\omega_{t}$ in Eq.\ref{eq:50}, the gCS term 
is (locally) $SO\left(2\right)_{L}$ gauge invariant, and accordingly
depends only on the metric, or, within the perturbative expansion,
on the symmetric part $H_{(ij)}$. From this expansion one can read
off the gCS contributions to the odd viscosity $\eta_{\text{o}}$ (Eq.\eqref{eq:18-1}), and to the odd, mixed, static susceptibility $\chi_{TJ,\text{o}}$ (Eq.\eqref{eq:17}).

\subsection{Additional terms at third order\label{subsec:Additional terms at third order}}

To obtain reliable results at $O\left(p^{3}\right)$ we, in principle,
need the full Lagrangian $\mathcal{L}_{3}$, which includes, but is
not equal to, $\mathcal{L}_{\text{gCS}}$. Nevertheless, we argue
that $\mathcal{L}_{3}-\mathcal{L}_{\text{gCS}}$ does not contribute
to the quantity of interest in this paper - $\eta_{\text{o}}$ to
$O\left(q^{2}\right)$. We already demonstrated in Appendix \ref{subsec:Odd-viscosity-from}
that the vertex part of the odd viscosity $\eta_{\mathcal{V}}$ is
independent of $\mathcal{L}_{3}$, and it remains to show that the
contact term part $\eta_{\mathcal{C}}$ is independent of $\mathcal{L}_{3}-\mathcal{L}_{\text{gCS}}$.
We do not have a general proof, but we address this issue in two ways: 
\begin{enumerate}
\item Within the microscopic model \eqref{eq:3-1-1}, the perturbative computation
of Appendix \ref{subsec:Perturbative-expansion} provides an explicit
expression for $\eta_{\mathcal{C}}$, which is completely saturated
by the effective action presented thus far. Thus $\eta_{\mathcal{C}}$
is independent of $\mathcal{L}_{3}-\mathcal{L}_{\text{gCS}}$ in the
particular realization \eqref{eq:3-1-1}. 
\item The term $\mathcal{L}_{3}$ is $P,T$-odd, and therefore vanishes
in an $s$-wave SF. On the other hand, it suffices to consider the
$g$s-wave SF where $s_{\theta}=0$ (but $\ell\neq0$), since for
$s_{\theta}\neq0$ the spin connection included in $\nabla_{\mu}\theta$
will only produce $O\left(p^{4}\right)$ corrections. By contracting
Galilean vectors, we were able to construct four $P,T$-odd terms
in $\mathcal{L}_{3}-\mathcal{L}_{\text{gCS}}$ for the $g$s-wave
SF, 
\begin{align}
\mathcal{L}_{3}-\mathcal{L}_{\text{gCS}}\supset & \ell\left[C_{1}\left(X\right)\tilde{E}_{i}E_{\omega}^{i}+C_{2}\left(X\right)\varepsilon^{ij}\tilde{E}_{i}E_{\omega,j}+C_{3}\left(X\right)\partial_{i}XE_{\omega}^{j}+C_{4}\left(X\right)\varepsilon^{ij}\partial_{i}XE_{\omega,j}\right].
\end{align}
where $\tilde{E}_{i}$ is the electric field of the improved $U\left(1\right)$
connection $\tilde{A}_{t}=A_{t}+\frac{1}{2m}\nabla^{i}\theta\nabla_{i}\theta,\;\tilde{A}_{i}=\partial_{i}\theta-s_{\theta}\omega_{i}$
\cite{hoyos2014effective}. Perturbatively expanding these, we do
not find any $O\left(q^{2}\right)$ contributions to $\eta_{\mathcal{C}}$
(or to $\eta_{\text{\ensuremath{\mathcal{V}}}}$, in accordance with
Appendix \ref{subsec:Odd-viscosity-from}).
\end{enumerate}

\section{Induced action \label{subsec:Obtaining--from}}

The arguments presented in the main text suffice to establish the
quantization of $\tilde{\eta}_{\text{o}}$ and $\tilde{\chi}_{TJ,\text{o}}$
directly from $S_{\text{eff}}$ - an explicit expression for $S_{\text{ind}}$
is not required. Nevertheless, it is instructive to compute certain
contributions in $S_{\text{ind}}$ to demonstrate these results 
explicitly, and also to reproduce simpler properties of $\ell$-wave
SFs. Here we will compute the contribution of $\mathcal{L}_{0}+\mathcal{L}_{2}^{\left(1\right)}\subset\mathcal{L}_{\text{eff}}$
to the induced Lagrangian $\mathcal{L}_{\text{ind}}$, and, along
the way, demonstrate explicitly the relation between $\text{var}s=0$
QH states and CSFs alluded to in the discussion Sec.\ref{sec:discussion}.

The starting point is the induced action due to $\mathcal{L}_{0}=P\left(X\right)$,
obtained from \eqref{eq:51}. It is given by 
\begin{align}
\mathcal{L}_{\text{ind}}= & -\frac{1}{2}\mathcal{V}\mathcal{G}\mathcal{V}+\mathcal{C}\label{eq:75}\\
= & P_{0}\sqrt{G}-n_{0}\mathcal{A}_{t}\nonumber \\
 & +\frac{1}{2}\frac{n_{0}}{m}\frac{\mathcal{B}^{2}-c_{s}^{-2}\mathcal{E}^{2}+\frac{s_{\theta}c_{s}^{-2}}{m}\mathcal{E}^{i}\partial_{i}B-\frac{s_{\theta}^{2}c_{s}^{-2}}{4m^{2}}\left(\partial B\right)^{2}}{\partial^{2}-c_{s}^{-2}\partial_{t}^{2}}\nonumber \\
 & -n_{0}\frac{m\left(\partial_{t}\sqrt{G}\right)^{2}/2+\left(\mathcal{E}^{i}-\frac{s_{\theta}}{2m}\partial_{i}B\right)\partial_{i}\sqrt{G}}{\partial^{2}-c_{s}^{-2}\partial_{t}^{2}}.\nonumber 
\end{align}
This expression contains, rather compactly, the entire linear response
of the $\ell$-wave SF to $O\left(p\right)$ in the derivative expansion,
as well as certain $O\left(p^{2}\right)$ contributions \cite{hoyos2014effective},
and should be interpreted as expanded to second order using \eqref{eq:48}-\eqref{eq:49}.
In using \eqref{eq:49}, one can set $H_{[AB]}=0$, since $S_{\text{ind}}$
is $SO\left(2\right)_{L}$ invariant and the anti-symmetric part $H_{[AB]}$
corresponds to the $SO\left(2\right)_{L}$ phase of the vielbein $E_{A}^{\;i}$.
Technically, $H_{[AB]}$ always appears in the combination $\partial_{\mu}\left(\theta+s_{\theta}\varepsilon^{AB}H_{AB}/2\right)\subset\nabla_{\mu}\theta$,
so that integrating out $\theta$ eliminates $H_{[AB]}$. 

Note that, diagrammatically, equation \eqref{eq:75} corresponds to
linear response at tree-level. Higher orders in $\theta$ will generate
diagrams with $\theta$ running in loops, which can be shown to produce
$O\left(p^{3}\right)$ corrections above the leading order to any
observable \cite{son2006general}, and are therefore irrelevant for
the purpose of $q^{2}$ corrections to $\eta_{\text{o}}$. 

The $O\left(p^{0}\right)$ part of \eqref{eq:75} is obtained by setting
$s_{\theta}=0$, as in an $s$-wave SF, 

\begin{align}
\mathcal{L}_{\text{ind},0}= & P_{0}\sqrt{G}-n_{0}A_{t}+\frac{1}{2}\frac{n_{0}}{m}\frac{B^{2}-c_{s}^{-2}E^{2}}{\partial^{2}-c_{s}^{-2}\partial_{t}^{2}}\\
 & -n_{0}\frac{m\left(\partial_{t}\sqrt{G}\right)^{2}/2+E^{i}\partial_{i}\sqrt{G}}{\partial^{2}-c_{s}^{-2}\partial_{t}^{2}}\nonumber 
\end{align}
The first line contains the ground state pressure and density $P_{0},n_{0}$,
as well as the London diamagnetic function $\rho_{\text{e}}=\frac{n_{0}}{m}\frac{1}{q^{2}-c_{s}^{-2}\omega^{2}}$
and the ideal Drude longitudinal conductivity $\sigma_{\text{e}}=-\frac{n_{0}}{m}\frac{i\omega c_{s}^{-2}}{q^{2}-c_{s}^{-2}\omega^{2}}$
of the SF \cite{hoyos2014effective}. The second line contains the
mixed response and mixed static susceptibility 
\begin{align}
\kappa_{\text{e}}^{ij,k} & =-n_{0}\delta^{ij}\frac{iq^{k}}{q^{2}-c_{s}^{-2}\omega^{2}},\\
\chi_{TJ,\text{e}}^{ij,t} & =n_{0}\delta^{ij}\frac{q^{2}}{q^{2}-c_{s}^{-2}\omega^{2}},\nonumber 
\end{align}
defined in Sec.\ref{sec: induced action and linear response}, as well as the inverse compressibility
$K^{-1}=-n_{0}m\frac{\omega^{2}}{q^{2}-c_{s}^{-2}\omega^{2}}$ (which
agrees with the thermodynamic expression $K^{-1}=n_{0}^{2}\frac{\partial\mu}{\partial n_{0}}=n_{0}mc_{s}^{2}$
at $q=0$). In particular, the $\ell$-wave SF is indeed a superfluid
- the even viscosity $\eta_{\text{e}}$ vanishes to zeroth order in
derivatives (see \cite{PhysRevB.86.245309} for a subtlety in separating
$K^{-1}$ from $\eta_{\text{e}}$). 

The $O\left(p\right)$ part of the \eqref{eq:75} is $P,T$-odd and
vanishes when $s_{\theta}=0$. It is given by 
\begin{align}
\mathcal{L}_{\text{ind},1}= & -s_{\theta}n_{0}\omega_{t}\label{eq:76-1-1}\\
 & +\frac{1}{2}\frac{s_{\theta}n_{0}}{m^{2}c_{s}^{2}}\frac{E^{i}\partial_{i}B}{\partial^{2}-c_{s}^{-2}\partial_{t}^{2}}\nonumber \\
 & -s_{\theta}n_{0}\frac{\left(E_{\omega}^{i}-\frac{1}{2m}\partial_{i}B\right)\partial_{i}\sqrt{G}}{\partial^{2}-c_{s}^{-2}\partial_{t}^{2}}.\nonumber 
\end{align}
The first and third lines produce the following odd viscosity \cite{hoyos2014effective},
\begin{align}
\eta_{\text{o}}^{\left(1\right)}= & -\frac{1}{2}s_{\theta}n_{0},\label{eq:14-1}\\
\eta_{\text{o}}^{\left(2\right)}= & \frac{1}{2}s_{\theta}n_{0}\frac{1}{q^{2}-c_{s}^{-2}\omega^{2}},\nonumber 
\end{align}
and setting $\omega=0$ one obtains the leading terms in Eq.\eqref{eq:14}.  By using the identity (up to a total derivative)
\begin{align}
 & E^{i}\partial_{i}B=\frac{1}{2}\varepsilon^{\mu\nu\rho}A_{\mu}\partial_{\nu}\partial^{2}A_{\rho},\label{eq:81}
\end{align}
 the second line of Eq.\eqref{eq:76-1-1} can be written as a non-local
CS term 

\begin{align}
\mathcal{L}_{\text{ind}}\supset & \frac{1}{2}\sigma_{\text{o}}\left(\omega,q\right)\varepsilon^{\mu\nu\rho}A_{\mu}ip_{\nu}A_{\rho}\label{eq:81-1}
\end{align}
with the odd (or Hall) conductivity $\sigma_{\text{o}}\left(\omega,q\right)=\sigma_{\text{o}}^{0}q^{2}/\left(q^{2}-c_{s}^{-2}\omega^{2}\right)$,
$\sigma_{\text{o}}^{0}=s_{\theta}n_{0}/2m^{2}c_{s}^{2}$ \cite{volovik1988quantized,goryo1998abelian,goryo1999observation,furusaki2001spontaneous,stone2004edge,roy2008collective,lutchyn2008gauge,hoyos2014effective},
with $\sigma_{\text{o}}\left(0,q\right)=\sigma_{\text{o}}^{0}$ unquantized,
and $\sigma_{\text{o}}\left(\omega,0\right)=0$, in accordance with
the boundary $U\left(1\right)_{N}$-neutrality \cite{PhysRevB.98.064503}.

To demonstrate explicitly that $c$ cannot be extracted from the odd
viscosity alone, it suffices to add the $O\left(p^{2}\right)$ term
$\mathcal{L}_{2}^{\left(1\right)}=F_{1}\left(X\right)R\subset\mathcal{L}_{2}$.
The situation is particularly simple for the special case $F_{1}\left(X\right)=-s_{\theta}^{2}P'\left(X\right)/4m$.
Then
\begin{align}
P\left(X-\frac{s_{\theta}^{2}}{4m}R\right) & =P\left(X\right)-\frac{s_{\theta}^{2}}{4m}P'\left(X\right)R+O\left(p^{4}\right)\\
 & =P\left(X\right)+F_{1}\left(X\right)R+O\left(p^{4}\right),\nonumber 
\end{align}
which shows that $F_{1}\left(X\right)R$ can be absorbed into $P\left(X\right)$
by a modification of $X$. The scalar $X-\frac{s_{\theta}^{2}}{4m}R$
is useful because, unlike $X$, it depends on $A_{\mu}$ and $\omega_{\mu}$
\textit{only} through the combination $\mathcal{A}_{\mu}=A_{\mu}+s_{\theta}\omega_{\mu}$.
This is evident in \eqref{eq:69}, where $B$ rather than $\mathcal{B}=B+\frac{s}{2}R$
appears. It is then clear that, to $O\left(p^{3}\right)$, adding
$\mathcal{L}_{2}^{\left(1\right)}=F_{1}\left(X\right)R=-\frac{s_{\theta}^{2}}{4m}P'\left(X\right)R$
to $\mathcal{L}_{0}=P\left(X\right)$ amounts to changing $B$ to
$\mathcal{B}$ in the induced Lagrangian \eqref{eq:75}, 
\begin{align}
\mathcal{L}_{\text{ind}}= & P_{0}\sqrt{G}-n_{0}\mathcal{A}_{t}\\
 & +\frac{1}{2}\frac{n_{0}}{m}\frac{\mathcal{B}^{2}-c_{s}^{-2}\mathcal{E}^{2}+\frac{s_{\theta}c_{s}^{-2}}{m}\mathcal{E}^{i}\partial_{i}\mathcal{B}-\frac{s_{\theta}^{2}c_{s}^{-2}}{4m^{2}}\left(\partial\mathcal{B}\right)^{2}}{\partial^{2}-c_{s}^{-2}\partial_{t}^{2}}\nonumber \\
 & -n_{0}\frac{m\left(\partial_{t}\sqrt{G}\right)^{2}/2+\left(\mathcal{E}^{i}-\frac{s_{\theta}}{2m}\partial_{i}\mathcal{B}\right)\partial_{i}\sqrt{G}}{\partial^{2}-c_{s}^{-2}\partial_{t}^{2}}.\nonumber 
\end{align}
The only contribution to $\eta_{\text{o}}$, beyond \eqref{eq:14-1},
comes from the term proportional to $\mathcal{E}^{i}\partial_{i}\mathcal{B}$.
By using the identity \eqref{eq:81} for $\mathcal{A}_{\mu}$, this
term can be written as the sum of non-local CS, WZ1, and WZ2 terms,
which generalizes \eqref{eq:81-1} to 
\begin{align}
\mathcal{L}_{\text{ind}}\supset & \frac{1}{2}\sigma_{\text{o}}\left(\omega,q\right)\varepsilon^{\mu\nu\rho}\left(A_{\mu}+s_{\theta}\omega_{\mu}\right)ip_{\nu}\left(A_{\rho}+s_{\theta}\omega_{\rho}\right).\label{eq:16-1-1}
\end{align}
Most importantly, this includes a non-local version of WZ2, which
is indistinguishable from $\mathcal{L}_{\text{gCS}}$ at $\omega=0$,
where $\sigma_{\text{o}}\left(0,q\right)=\sigma_{\text{o}}^{0}$ is
a constant. Noting that $F_{1}'=-s_{\theta}^{2}P''/4m=-\left(s_{\theta}/2\right)\sigma_{\text{o}}^{0}$,
and comparing to \eqref{eq:54-2}, it follows that $c$ and $F_{1}'$
will enter the $\omega=0$ odd viscosity only through the combination
$c+48\pi s_{\theta}F_{1}'$. In more detail, the odd viscosity tensor
due to $\mathcal{L}_{0}+\mathcal{L}_{2}^{\left(1\right)}+\mathcal{L}_{\text{gCS}}$,
is given by 

\begin{align}
\eta_{H}^{\left(1\right)}\left(\omega,q^{2}\right)= & -\frac{1}{2}s_{\theta}n_{0}-\left(\frac{c}{24}\frac{1}{4\pi}+\frac{s_{\theta}}{2}F_{1}'\frac{q^{2}}{q^{2}-c_{s}^{-2}\omega^{2}}\right)q^{2}+O\left(q^{4}\right),\label{eq:86}\\
\eta_{H}^{\left(2\right)}\left(\omega,q^{2}\right)= & \frac{1}{2}s_{\theta}n_{0}\frac{1}{q^{2}-c_{s}^{-2}\omega^{2}}+\left(\frac{c}{24}\frac{1}{4\pi}+\frac{s_{\theta}}{2}F_{1}'\frac{q^{2}}{q^{2}-c_{s}^{-2}\omega^{2}}\right)+O\left(q^{2}\right),\nonumber 
\end{align}
which, at $\omega=0$, is a special case of equation (12) of the main
text. 

Equation \eqref{eq:86} remains valid away from the special point
$F_{1}=-s_{\theta}^{2}P'/4m$, even though \eqref{eq:16-1-1} does
not. Examining the perturbatively expanded $\mathcal{L}_{0}$ \eqref{eq:51}
and $\mathcal{L}_{2}^{\left(1\right)}$ \eqref{eq:54}, we see that
a general $F_{1}$ amounts to replacing $B$ in \eqref{eq:75} with
$B+\alpha\frac{s_{\theta}}{2}R$, where $\alpha=-\frac{4mF_{1}'}{s_{\theta}^{2}P''}\neq1$
generically (as well as in the microscopic model \eqref{eq:c1-1}).
The general induced Lagrangian due to $\mathcal{L}_{0}+\mathcal{L}_{2}^{\left(1\right)}$,
valid to $O\left(p^{3}\right)$, is then given by
\begin{align}
\mathcal{L}_{\text{ind}}= & P_{0}\sqrt{G}-n_{0}\mathcal{A}_{t}\label{eq:87}\\
 & +\frac{1}{2}\frac{n_{0}}{m}\frac{\mathcal{B}^{2}-c_{s}^{-2}\mathcal{E}^{2}+\frac{s_{\theta}c_{s}^{-2}}{m}\mathcal{E}^{i}\partial_{i}\left(B+\alpha\frac{s_{\theta}}{2}R\right)-\frac{s_{\theta}^{2}c_{s}^{-2}}{4m^{2}}\left(\partial B\right)^{2}}{\partial^{2}-c_{s}^{-2}\partial_{t}^{2}}\nonumber \\
 & -n_{0}\frac{m\left(\partial_{t}\sqrt{G}\right)^{2}/2+\left[\mathcal{E}^{i}-\frac{s_{\theta}}{2m}\partial_{i}\left(B+\alpha\frac{s_{\theta}}{2}R\right)\right]\partial_{i}\sqrt{G}}{\partial^{2}-c_{s}^{-2}\partial_{t}^{2}},\nonumber 
\end{align}
and, along with the $\mathcal{L}_{\text{gCS}}$ , produces the odd
viscosity \eqref{eq:86}. This expression does not depend on $A_{\mu},\omega_{\mu}$
only through $\mathcal{A}_{\mu}$, but the terms contributing to \eqref{eq:86}
still vanish  $s_{\theta}=0$, which is why the \textit{improved}
odd viscosity due to \eqref{eq:87} vanishes. In addition to $\mathcal{L}_{2}^{\left(1\right)}$,
the second order terms $\mathcal{L}_{2}^{\left(2\right)},\mathcal{L}_{2}^{\left(3\right)}$
\eqref{eq:49-2} also produce $q^{2}$ corrections to the odd viscosity,
but not to the improved odd viscosity.

Though Eq.\eqref{eq:16-1-1} describes only a part of $\mathcal{L}_{\text{ind}}$,
and is non-generic, it does reveal the analogy between 
CSFs and $\text{var}s=0$ QH states, described in the discussion Sec.\ref{sec:discussion}
 in a very simple setting. Indeed, comparing Eq.\eqref{eq:16-1-1}
with Eq.\eqref{eq:11} we see that CSFs
are analogous to $\text{var}s=0$ QH states, with $\overline{s}=-s_{\theta}=-\ell/2$,
but with a non-local, non-quantized, Hall conductivity, in place of
the filling factor $\nu/2\pi$. Additionally, both QH states and CSFs have the same gCS term Eq.\eqref{eq:54-2}, with $c$ the boundary
chiral central charge.

\section{Detailed analysis of the microscopic model \eqref{eq:3-1-1}\label{sec:microscopic model } }



\subsection{Symmetry \label{subsec:Symmetry}}

The action $S_{\text{m}}$ is invariant under $U\left(1\right)_{N}$ gauge transformations,
\begin{align}
 & \psi\mapsto e^{-i\alpha}\psi,\;\Delta^{j}\mapsto e^{-2i\alpha}\Delta^{j},\;A_{\mu}\mapsto A_{\mu}+\partial_{\mu}\alpha,
\end{align}
which implies the current conservation $\partial_{\mu}(\sqrt{G}J^{\mu})=0$,
where $\sqrt{G}J^{\mu}=-\delta S/\delta A_{\mu}$. It is also clear
that $S_{\text{m}}$ is invariant under \textit{time-independent} spatial diffeomorphisms,
generated by $\delta x^{i}=\xi^{i}\left(\mathbf{x}\right)$, if $\psi$
transforms as a function, $A_{\mu}$ as a 1-form, $\Delta^{j}$ as
a vector, and $G_{ij}$ as a rank-2 tensor. As described in section \ref{sec: building blocks}, due to its Galilean symmetry in flat space, $S_{\text{m}}$ is also invariant
under time-dependent spatial diffeomorphisms $\delta x^{i}=\xi^{i}\left(\mathbf{x},t\right)$,
provided one modifies the transformation rule of $A_{i}$ to Eq.\eqref{eq:4-3-1-1}.

\subsection{Effective action and fermionic Green's function }

Starting with the microscopic action \eqref{eq:3-1-1}, the effective
action for the order parameter $\Delta$ in the $A,G$ background
is obtained by integrating out the (generically) gapped fermion $\psi$,
\begin{align}
 & e^{iS_{\text{eff,m}}\left[\Delta;A,G\right]}=\int\text{D}\left(G^{1/4}\psi\right)\text{D}\left(G^{1/4}\psi^{\dagger}\right)e^{iS_{\text{m}}\left[\psi;\Delta,A,G\right]},\label{eq:20}
\end{align}
where $G^{1/4}=\left(\text{det}G_{ij}\right)^{1/4}$ is the square
root of the volume element $\sqrt{G}$. The form of the measure is
fixed by the fact that the fundamental fermionic degree of freedom
is the fermion-density $\tilde{\psi}=G^{1/4}\psi$, which satisfies
the usual canonical commutation relation $\left\{ \tilde{\psi}^{\dagger}\left(\mathbf{x}\right),\tilde{\psi}\left(\mathbf{y}\right)\right\} =\delta^{\left(2\right)}\left(\mathbf{x}-\mathbf{y}\right)$
as an operator \cite{hawking1977zeta,fujikawa1980comment,abanov2014electromagnetic,PhysRevB.98.064503}.
This is to be contrasted with $\left\{ \psi^{\dagger}\left(\mathbf{x}\right),\psi\left(\mathbf{y}\right)\right\} =\delta^{\left(2\right)}\left(\mathbf{x}-\mathbf{y}\right)/\sqrt{G\left(\mathbf{x}\right)}$
which ties the fermion to the background metric. 

In terms of $\tilde{\psi}$ the action \eqref{eq:3-1-1} takes the form
\begin{align}
S_{\text{m}} & =\int\text{d}^{2}x\text{d}t\left[\tilde{\psi}^{\dagger}\frac{i}{2}\overleftrightarrow{\nabla}_{t}\tilde{\psi}-\frac{1}{2m}G^{ij}\nabla_{i}\tilde{\psi}^{\dagger}\nabla_{j}\tilde{\psi}+\left(\frac{1}{2}\Delta^{i}\tilde{\psi}^{\dagger}\nabla_{i}\tilde{\psi}^{\dagger}+h.c\right)-\mathcal{U}\right],\label{eq:78}
\end{align}
where $\nabla_{\mu}=\partial_{\mu}+iA_{\mu}-\frac{1}{4}\partial_{\mu}\log G$
is the covariant derivative for densities, and $\mathcal{U}=\frac{1}{2\lambda}\sqrt{G}G_{ij}\Delta^{i*}\Delta^{j}$.
  Passing to the BdG form of the fermionic part of the action,
in terms of the Nambu spinor-density $\tilde{\Psi}^{\dagger}=\left(\tilde{\psi}^{\dagger},\tilde{\psi}\right)$
(which is a Majorana spinor-density \cite{PhysRevB.98.064503}), one
finds 
\begin{align}
S_{\text{m}}=\int\text{d}^{2}x\text{d}t & \left\{ \frac{1}{2}\tilde{\Psi}^{\dagger}\gamma^{0}\left[i\gamma^{0}\partial_{t}-A_{t}+\frac{1}{2m}\nabla_{i}G^{ij}\nabla_{j}\right.\right.\label{eq:22}\\
 & +\left.\left.\frac{i}{2}\gamma^{\tilde{A}}\left(e_{\tilde{A}}^{\;i}\partial_{i}+\partial_{i}e_{\tilde{A}}^{\;i}\right)\right]\tilde{\Psi}-\mathcal{U}\right\} \nonumber \\
=\int\text{d}^{2}x\text{d}t & \left\{ \frac{1}{2}\tilde{\Psi}^{\dagger}\gamma^{0}\mathcal{G}^{-1}\tilde{\Psi}-\mathcal{U}\right\} ,\nonumber 
\end{align}
where derivatives act on all fields to the right; $\tilde{A}=1,2$
is an index for $U\left(1\right)_{N}$, viewed as a copy of $SO\left(2\right)$;
the gamma matrices are $\gamma^{0}=\sigma^{z},\;\gamma^{1}=-i\sigma^{x},\;\gamma^{2}=i\sigma^{y}$,
satisfying $\left\{ \gamma^{\mu},\gamma^{\nu}\right\} =2\eta^{\mu\nu}$
with $\eta^{\mu\nu}=\text{diag}\left[1,-1,-1\right]$, and $\text{tr}\left(\gamma^{0}\gamma^{1}\gamma^{2}\right)=2i$;
and 
\begin{align}
 & e_{\tilde{A}}^{\;i}=\begin{pmatrix}\text{Re}\Delta^{x} & \text{Re}\Delta^{y}\\
\text{Im}\Delta^{x} & \text{Im}\Delta^{y}
\end{pmatrix}
\end{align}
is the \textit{emergent} vielbein \cite{volovik1990gravitational,PhysRevB.98.064503},
to be distinguished from the background vielbein $E_{A}^{\;i}$ (with
an $SO\left(2\right)_{L}$ index $A=1,2$) that appeared in the main
text and that will be used momentarily. We also defined the inverse
Green's function $\mathcal{G}^{-1}$. The effective action \eqref{eq:20}
is then given by the logarithm of the Pfaffian 
\begin{align}
S_{\text{eff,m}}= & -i\log\text{Pf}\left(i\gamma^{0}\mathcal{G}^{-1}\right)-\int\text{d}^{2}x\text{d}t\mathcal{U}\label{eq:24-1}\\
= & -\frac{i}{2}\log\text{Det}\left(i\gamma^{0}\mathcal{G}^{-1}\right)-\int\text{d}^{2}x\text{d}t\mathcal{U}.\nonumber 
\end{align}

\subsection{Fermionic ground state topology }

For a given $\Delta^{j}$, the fermion $\psi$ is gapped, unless the
chemical potential $\mu$ or chirality $\ell=\text{sgn}\left(\text{Im}\left(\Delta^{x*}\Delta^{y}\right)\right)$
 are tuned to 0, and forms a fermionic topological phase characterized
by the bulk Chern number.  Assuming $A_{\mu}=0$ and space-time
independent $\Delta^{i},G^{ij}$, it is given by \cite{volovik2009universe}
\begin{align}
C= & \frac{1}{24\pi^{2}}\mbox{tr}\int\mbox{d}^{3}q\varepsilon^{\alpha\beta\gamma}\left(\mathcal{G}\partial_{\alpha}\mathcal{G}^{-1}\right)\left(\mathcal{G}\partial_{\beta}\mathcal{G}^{-1}\right)\left(\mathcal{G}\partial_{\gamma}\mathcal{G}^{-1}\right)\in\mathbb{Z},\label{eq:28-1}
\end{align}
and determines the boundary chiral central charge $c=C/2$ \cite{read2000paired,kitaev2006anyons,volovik2009universe,ryu2010topological}.
Here the fermionic Green's function $\mathcal{G}$ is Fourier transformed
to Euclidian 3-momentum $q=\left(iq_{0},\mathbf{q}\right)$ (see \eqref{eq:184-1}).
For the particular model \eqref{eq:3-1-1} one finds 
\begin{align}
 & c=-\left(\ell/4\right)\left(\text{sgn}\left(\mu\right)+\text{sgn}\left(m\right)\right)\in\left\{ 0,\pm1/2\right\} ,\label{eq:29}
\end{align}
see \cite{read2000paired,volovik2009universe,PhysRevB.98.064503}
for similar expressions. Note that the central charge is well defined
for both $m>0$ and $m<0$, even though the single particle dispersion
is not bounded from below in the latter, and many physical quantities
naively diverge (we will see below that certain physical quantities
diverge also with $m>0$). A negative mass can occur as an effective
mass in lattice models, in which case the lattice spacing provides
a natural cutoff (which must be smooth in momentum space for \eqref{eq:28-1}
to hold). In any case, a negative mass makes it possible to obtain
both fundamental central charges $c=\pm1/2$, for fixed $\ell$, within
the model \eqref{eq:3-1-1}. All possible $c\in\left(1/2\right)\mathbb{Z}$
can then be obtained by stacking layers of the model \eqref{eq:3-1-1}
with the same $\ell$ but different $m,\mu$. Thus the model \eqref{eq:3-1-1}
suffices to generate a representative for all topological phases of
the $p$-wave CSF. For concreteness, below we will work only with $m>0$,
in which case $c$ is given by Eq.\eqref{eq:12-1}.

\subsection{Symmetry breaking and bosonic ground state in the presence of a background
metric\label{subsec:Symmetry-breaking-and}}

For time independent fields $A,G,\Delta$ the effective action reduces
to 
\begin{align}
S_{\text{eff,m}}\left[\Delta;G\right] =&-\int\text{d}^{2}x\text{d}t\varepsilon_{0}\left[\Delta;G\right],
\end{align}
where $\varepsilon_{0}$ is the ground-state energy-density as a function
of the fields. In flat space $G_{ij}=\delta_{ij}$, with $A_{t}=-\mu$
and $A_{i}=0$, and assuming $\Delta$ is constant, it is given by
\cite{volovik2009universe,PhysRevB.98.064503} 
\begin{align}
\varepsilon_{0}= & \frac{1}{2}\int\frac{\text{d}^{2}\mathbf{q}}{\left(2\pi\right)^{2}}\left(\xi_{\mathbf{q}}-\sqrt{\xi_{\mathbf{q}}^{2}+g^{ij}q_{i}q_{j}}\right)+\frac{1}{2\lambda}\delta_{ij}g^{ij},\label{eq:27}
\end{align}
where 
\begin{align}
 & \xi_{\mathbf{q}}=\left|\mathbf{q}\right|^{2}/2m-\mu\label{eq:33}
\end{align}
is the single particle dispersion, and $g^{ij}=\Delta^{(i}\Delta^{j)*}=\delta^{\tilde{A}\tilde{B}}e_{\tilde{A}}^{\;i}e_{\tilde{B}}^{\;j}$
is the \textit{emergent metric} - a dynamical metric to be distinguished
from the background metric $G^{ij}$. The ground state configuration
of $g^{ij}$ is determined by minimizing $\varepsilon_{0}$, while
the overall phase $\theta$ of the order parameter and the chirality
$\ell$, of which $g^{ij}$ is independent, are left undetermined.
Thus $g^{ij}$ corresponds to a massive Higgs field, while $\theta$
is a Goldstone field. The energy-density \eqref{eq:27} is UV divergent,
and requires regularization. We do this in the simplest manner, by
introducing a momentum cutoff $q^{2}<\Lambda^{2}$. Since the divergence
disappears for $g^{ij}=0$ (assuming $m>0$), this can be thought
of as a small, but non-vanishing, range $1/\Lambda$ for the interaction
mediated by $\Delta$. With a finite $\Lambda$, the energy-density
is well defined and has a unique global minimum at $g^{ij}=\Delta_{0}^{2}\delta^{ij}$,
with $\Delta_{0}$ determined by the self-consistent equation 
\begin{align}
 & \frac{1}{4}\int^{\Lambda}\frac{\text{d}^{2}\boldsymbol{q}}{\left(2\pi\right)^{2}}\frac{\left|\boldsymbol{q}\right|^{2}}{\sqrt{\xi_{\boldsymbol{q}}^{2}+\Delta_{0}^{2}\left|\boldsymbol{q}\right|^{2}}}=\frac{1}{\lambda}.\label{eq:28}
\end{align}
For $\mu>0$ the non-interacting system has a Fermi surface, and a
solution exists for all $\lambda>0$, which is the statement of the
BCS instability. For $\mu<0$, the non-interacting system is gapped,
and a solution exists if the interaction is large enough compared
with the gap, $\lambda\Lambda^{-4}\gtrsim\left|\mu\right|$.

Consider now the case of a general constant metric $G_{ij}$, and
let us introduce a constant vielbein $E$ such that $G_{ij}=E_{\;i}^{A}\delta_{AB}E_{\;j}^{B}$.
The inverse transpose $E^{-T}=\left(E^{-1}\right)^{T}$ is given in
coordinates by $E_{A}^{\;i}$. We also introduce the internal order
parameter $\Delta^{A}=E_{\;i}^{A}\Delta^{i}$. The action \eqref{eq:78}
then reduces to 
\begin{align}
S_{\text{m}} & =\int\text{d}^{2}x\text{d}t\left[\tilde{\psi}^{\dagger}i\partial_{t}\tilde{\psi}-\frac{\delta^{AB}}{2m}E_{A}^{\;i}\partial_{i}\tilde{\psi}^{\dagger}E_{B}^{\;j}\partial_{j}\tilde{\psi}+\left(\frac{1}{2}\Delta^{A}E_{A}^{\;i}\tilde{\psi}^{\dagger}\partial_{i}\tilde{\psi}^{\dagger}+h.c\right)-\frac{1}{2\lambda}\delta_{AB}\Delta^{A*}\Delta^{B}\right].\label{eq:78-1}
\end{align}
This is identical to the flat space case, with $\partial_{i}$ replaced
by $E_{A}^{\;i}\partial_{i}$. We also need to change the UV cutoff
to $\delta^{AB}E_{A}^{\;i}q_{i}E_{B}^{\;j}q_{j}=G^{ij}q_{i}q_{j}<\Lambda^{2}$.
This in natural since we interpret $\Lambda^{2}$ as a range of the
interaction mediated by $\Delta$, which should be defined in terms
of the geodesic distance rather than the Euclidian distance. It follows
that the flat space result \eqref{eq:27} is modified to 
\begin{align}
\varepsilon_{0} & =\frac{1}{2}\int_{\left|E^{-T}\mathbf{q}\right|^{2}<\Lambda^{2}}\frac{\text{d}^{2}\mathbf{q}}{\left(2\pi\right)^{2}}\left(\xi_{E^{-T}\mathbf{q}}-\sqrt{\xi_{E^{-T}\mathbf{q}}^{2}+g^{AB}E_{A}^{\;i}E_{B}^{\;j}q_{i}q_{j}}\right)+\frac{1}{2\lambda}\delta_{AB}g^{AB}\label{eq:30}\\
 & =\frac{1}{2}\sqrt{G}\int_{q^{2}<\Lambda^{2}}\frac{\text{d}^{2}\mathbf{k}}{\left(2\pi\right)^{2}}\left(\xi_{\mathbf{k}}-\sqrt{\xi_{\mathbf{k}}^{2}+g^{AB}k_{A}k_{B}}\right)+\frac{1}{2\lambda}\delta_{AB}g^{AB},\nonumber 
\end{align}
where $\mathbf{k}=E^{-T}\mathbf{q}$, or $k_{A}=E_{A}^{\;i}q_{i}$,
and $g^{AB}=\Delta^{(A}\Delta^{B)*}=\delta^{\tilde{A}\tilde{B}}e_{\tilde{A}}^{\;A}e_{\tilde{B}}^{\;B}$
is the \textit{internal} emergent metric. This is identical to the
$G_{ij}=\delta_{ij}$ result \eqref{eq:27}, apart from the volume
element $\sqrt{G}$, and the fact that it is the internal metric $g^{AB}$
that appears, rather than $g^{ij}$. It is then clear that minimizing
\eqref{eq:30} with respect to $g^{AB}$ gives 
\begin{align}
 & g^{AB}=\Delta_{0}^{2}\delta^{AB}\text{, or }g^{ij}=\Delta_{0}^{2}G^{ij},
\end{align}
with the same $\Delta_{0}$ of \eqref{eq:28}, which is $G$ independent.
Thus, the emergent metric is proportional to the background metric
in the ground state.  This solution corresponds to emergent vielbeins
$e_{\tilde{A}}^{\;A}\in O\left(2\right)$, or order parameters $\Delta^{A}=\Delta_{0}e^{2i\theta}\left(1,\pm i\right)$,
which is the $p_{x}\pm ip_{y}$ configuration, and implies the SSB
pattern 
\begin{align}
 & \left(\mathbb{Z}_{2,T}\ltimes U\left(1\right)_{N}\right)\times\left(\mathbb{Z}_{2,P}\ltimes SO\left(2\right)_{L}\right)\rightarrow\begin{cases}
\mathbb{Z}_{2,PT}\ltimes U\left(1\right)_{L-\frac{\ell}{2}N} & \ell\in2\mathbb{Z}+1\\
\mathbb{Z}_{2,PT}\ltimes U\left(1\right)_{L-\frac{\ell}{2}N}\times\mathbb{Z}_{2,\left(-1\right)^{N}} & \ell\in2\mathbb{Z}
\end{cases},\label{eq:4-2-1-1}
\end{align}
described less formally in the main text. Note that fermion parity $\mathbb{Z}_{2,\left(-1\right)^{N}}$
is the $\mathbb{Z}_{2}$ subgroup of $U\left(1\right)_{L-\frac{\ell}{2}N}$
for odd $\ell$. For $\Delta^{j}$, we find the ground state configuration \eqref{eq:12}
- a result that was stated previously in the literature \cite{read2000paired,hoyos2014effective,moroz2015effective,moroz2016chiral,quelle2016edge},
and is derived here to zeroth order in derivatives.  

As described in Sec.\ref{sec:microscopic model }, we will ignore the massive Higgs fluctuations, and obtain $S_{\text{eff}}\left[\theta;A,G\right]$ by plugging the ground state configuration \eqref{eq:12} into the functional Pfaffian \eqref{eq:24-1}.

\subsection{Perturbative expansion\label{subsec:Perturbative-expansion}}

We now write $E_{A}^{\;i}=\delta_{A}^{i}+H_{A}^{\;i}$ and $e_{\tilde{A}}^{\;A}=\Delta_{0}\delta_{\tilde{A}}^{A}$
(which corresponds to $\Delta^{A}=\Delta_{0}\left(1,i\right)^{A}$)
and expand \eqref{eq:22} to second order in $H,A$. Due to $SO\left(2\right)_{L}$
gauge symmetry, the anti-symmetric part of $H$ can be interpreted
as the Goldstone field, $\theta=\left(s_{\theta}/2\right)\varepsilon_{AB}H^{AB}$,
as explained in Appendix \ref{subsec:Obtaining--from}. The $p_{x}-ip_{y}$
configuration $\Delta^{A}=\Delta_{0}\left(1,-i\right)^{A}$ can be
incorporated by changing the sign of one of the gamma matrices $\gamma^{\tilde{A}}$.
The expansion in $H,A$ produces a splitting of the the propagator
into an unperturbed propagator and vertices, $\mathcal{G}^{-1}=\mathcal{G}_{0}^{-1}+\mathcal{V}$,
where $\mathcal{V}$ further splits as $\mathcal{V}=\mathcal{V}_{1}+\mathcal{V}_{2}$,
where $\mathcal{V}_{1}$ ($\mathcal{V}_{2}$) is first (second) order
in the fields. The terms in $\mathcal{V}_{2}$ are often referred
to as contact terms. Using \eqref{eq:48} we find the explicit form
of $\mathcal{G}_{0}^{-1},\mathcal{V}_{1},\mathcal{V}_{2}$ in Fourier
components,
\begin{align}
\mathcal{G}_{0}^{-1}\left(q\right)= & -\gamma^{0}q_{0}-\Delta_{0}\gamma^{j}q_{j}-\xi_{\mathbf{q}},\label{eq:184-1}\\
\mathcal{V}_{1}\left(q,p\right)= & -A_{t,p}-\Delta_{0}\gamma^{A}\left(H_{A}^{\;i}\right)_{p}q_{i}\nonumber \\
 & -\frac{1}{m}\left[q_{i}q_{j}-\frac{1}{4}\left(p_{i}p_{j}-\delta_{ij}p^{2}\right)\right]H_{p}^{ij}+\gamma^{0}\frac{1}{m}A_{p}^{j}q_{j},\nonumber \\
\mathcal{V}_{2}\left(q,0\right)= & -\frac{1}{2m}\left(H_{A}^{\;i}H^{Aj}\right)_{p=0}q_{i}q_{j}-\frac{1}{8m}\left(\partial^{j}H_{A}^{\;A}\partial_{j}H_{B}^{\;B}\right)_{p=0}\nonumber \\
 & -\gamma^{0}\frac{2}{m}\left(A_{i}H^{(ij)}\right)_{p=0}q_{j}-\frac{1}{2m}\left(A^{j}A_{j}\right)_{p=0}.\nonumber 
\end{align}
Here $\left(\cdots\right)_{p}$ denotes the $p$ Fourier component
of the field $\left(\cdots\right)$, and we set $p=0$ in $\mathcal{V}_{2}$
since only this component will be relevant. The unperturbed Greens's
function is given explicitly by 
\begin{align}
 & \mathcal{G}_{0}\left(q\right)=-\frac{q_{0}\gamma^{0}+\Delta_{0}q_{i}\gamma^{i}-\xi_{\mathbf{q}}}{q_{0}^{2}-q_{i}q^{i}-\xi{}_{\mathbf{q}}^{2}}.
\end{align}
The perturbative expansion of $S_{\text{eff}}$ is obtained from \eqref{eq:24-1}
by using $\text{log}\left[\text{Det}\left(\cdot\right)\right]=\text{Tr}\left[\log\left(\cdot\right)\right]$,
and expanding the logarithm in $\mathcal{V}$,
\begin{eqnarray}
 & S_{\text{eff,m}} & =-i\text{Tr}\left\{ \log\left[i\gamma^{0}\left(\mathcal{G}_{0}^{-1}+\mathcal{V}\right)\right]\right\} \label{eq:25}\\
 &  & =-\frac{i}{2}\mbox{Tr}\left(\log i\gamma^{0}\mathcal{G}_{0}^{-1}\right)-\frac{i}{2}\mbox{Tr}\left(\mathcal{G}_{0}\mathcal{V}\right)+\frac{i}{4}\mbox{Tr}\left(\mathcal{G}_{0}\mathcal{V}\right)^{2}+O\left(\mathcal{V}^{3}\right)\nonumber \\
 &  & =-\frac{i}{2}\text{Tr}\left(\mathcal{G}_{0}\mathcal{V}_{1}\right)-\frac{i}{2}\text{Tr}\left(\mathcal{G}_{0}\mathcal{V}_{2}\right)+\frac{i}{4}\text{Tr}\left(\mathcal{G}_{0}\mathcal{V}_{1}\mathcal{G}_{0}\mathcal{V}_{1}\right)+\cdots,\nonumber 
\end{eqnarray}
where in the last line we kept explicit only terms at first and second
order in $H,A$ (the term of zeroth order was described in the previous
section). Writing the functional traces as integrals over Fourier
components and traces over spinor indices, we then find 
\begin{align}
S_{\text{eff,m}}= & -\frac{i}{2}\text{tr}\int_{q}\mathcal{V}_{1}\left(q,0\right)\mathcal{G}_{0}\left(q\right)-\frac{i}{2}\text{tr}\int_{q}\mathcal{V}_{2}\left(q,0\right)\mathcal{G}_{0}\left(q\right)\label{eq:39}\\
 & +\frac{i}{4}\text{tr}\int_{p,q}\mathcal{G}_{0}\left(q-\frac{1}{2}p\right)\mathcal{V}_{1}\left(q,-p\right)\mathcal{G}_{0}\left(q+\frac{1}{2}p\right)\mathcal{V}_{1}\left(q,p\right)+\cdots,\nonumber 
\end{align}
where $\int_{q}=\int\frac{\text{d}^{2}q\text{d}q_{0}}{\left(2\pi\right)^{3}}$.
We are interested in $S_{\text{eff}}$ to third order in derivatives,
which amounts to expanding the above expression to $O\left(p^{3}\right)$,
and evaluating the resulting traces and integrals. These computations
are performed in the accompanying Mathematica notebook. 

The result, focusing on terms relevant for $\eta_{\text{o}},\tilde{\eta}_{\text{o}}$
to $O\left(q^{2}\right)$, is compatible with the general effective
action of Sec.\ref{sec: effective field theory} and Appendix \ref{subsec:Effective-action-and},
as confirmed by comparing \eqref{eq:39} to the perturbatively expanded
$S_{\text{eff}}$.  This comparison provides explicit expressions
for all of the coefficients that appear in $S_{\text{eff}}$, as we
now describe. The ground state pressure $P\left(\mu\right)$ diverges
logarithmically, and is given by
\begin{align}
P & =\frac{1}{2}\int^{\Lambda}\frac{\text{d}^{2}q}{\left(2\pi\right)^{2}}\left[\frac{q^{2}}{2m}-\frac{\frac{1}{2}\Delta_{0}^{2}q^{2}+\frac{q^{2}}{2m}\left(\frac{q^{2}}{2m}-\mu\right)}{\sqrt{\Delta_{0}^{2}q^{2}+\left(\frac{q^{2}}{2m}-\mu\right)^{2}}}\right]\\
 & =-\frac{m^{3}\Delta_{0}^{4}}{4\pi}\left(1-2\frac{\mu}{m\Delta_{0}^{2}}\right)\log\Lambda+O\left(\Lambda^{0}\right).\nonumber 
\end{align}
Directly computing the ground state density $n_{0}$ and leading odd
viscosity $\eta_{\text{o}}^{\left(1\right)}$ one finds 
\begin{align}
n_{0} & =\frac{1}{2}\int\frac{\text{d}^{2}q}{\left(2\pi\right)^{2}}\left[1-\frac{\left(\frac{q^{2}}{2m}-\mu\right)}{\sqrt{\Delta_{0}^{2}q^{2}+\left(\frac{q^{2}}{2m}-\mu\right)^{2}}}\right]\\
 & =\frac{m^{2}\Delta_{0}^{2}}{2\pi}\log\Lambda+O\left(\Lambda^{0}\right),\nonumber \\
\eta_{\text{o}}^{\left(1\right)} & =-\frac{\ell}{16}\int\frac{\text{d}^{2}q}{\left(2\pi\right)^{2}}\frac{\Delta_{0}^{2}q^{2}\left(\frac{q^{2}}{2m}+\mu\right)}{\left[\left(\frac{q^{2}}{2m}-\mu\right)^{2}+q^{2}\Delta_{0}^{2}\right]^{3/2}}\\
 & =-\frac{\ell m^{2}\text{\ensuremath{\Delta}}_{0}^{2}}{8\pi}\log\Lambda+O\left(\Lambda^{0}\right),\nonumber 
\end{align}
so the relations $n_{0}=P'\left(\mu\right)$, and $\eta_{\text{o}}^{\left(1\right)}=-\left(\ell/4\right)n_{0}$,
described in the main text, are maintained to leading order in the
cutoff. 

As explained in Appendix \ref{subsec:Symmetry-breaking-and}, the cutoff
$\Lambda$ corresponds to a non-vanishing interaction range, which
softens the contact interaction in the model \eqref{eq:3-1-1}. With
a space-independent metric, a smooth cutoff can easily be implemented
by replacing 
\begin{align}
 & \Delta^{A}E_{A}^{\;j}\tilde{\psi}_{-\mathbf{q}}^{\dagger}iq_{j}\tilde{\psi}_{\mathbf{q}}^{\dagger}\mapsto\Delta^{A}E_{A}^{\;j}\tilde{\psi}_{-\mathbf{q}}^{\dagger}\left(iq_{j}e^{-q_{k}q_{l}G^{kl}/\Lambda^{2}}\right)\tilde{\psi}_{\mathbf{q}}^{\dagger},
\end{align}
for example, in the Fourier transformed Eq.\eqref{eq:78-1}, and should
lead to the \textit{exact }relations $n_{0}=P'\left(\mu\right)$,
$\eta_{\text{o}}^{\left(1\right)}=-\left(\ell/4\right)n_{0}$. However,
a computation of the $q^{2}$ correction to $\eta_{\text{o}}$ requires
a space-dependent metric, where a non-vanishing interaction range
involves the geodesic distance and complicates the vertex $\mathcal{V}$
in \eqref{eq:184-1} considerably. Moreover, all other coefficients
in $S_{\text{eff}}$ converge, and we can therefore work with the
simple contact interaction, $\Lambda=\infty$.

The coefficients $P'',F_{1}',F_{2},F_{3}$ were presented in Sec.\ref{sec:microscopic model }. The remaining coefficients $F_{4},F_{5},F_{6}$, are irrelevant for
the quantities discussed in the main text,  and are presented here
for completeness,
\begin{align}
F_{4}= & \frac{1}{24\pi\mu}\begin{cases}
\frac{\kappa-2}{2}\\
\frac{1}{1+2\kappa}
\end{cases},\;F_{5}=\frac{1}{24\pi\mu\Delta_{0}^{2}}\begin{cases}
1\\
-\frac{1}{\left(1+2\kappa\right)^{2}}
\end{cases},\;F_{6}=-\frac{\kappa}{24\pi\mu}\begin{cases}
\frac{1}{2}\\
\frac{1}{\left(1+2\kappa\right)^{2}}
\end{cases}.\label{eq:c3}
\end{align}

As stated in Sec.\ref{sec:microscopic model },  there is a sense in which the relativistic limit  $\kappa\rightarrow0$, or $m\rightarrow\infty$ reproduces the  effective action of a massive Majorana spinor
in Riemann-Cartan space-time \cite{PhysRevB.98.064503,hughes2013torsional}. In particular, in the limit $\kappa\rightarrow0$
the dimensionless coefficients \eqref{eq:c1-1} are all quantized, as
follows from dimensional analysis. Apart from $c$, only the coefficient
$F_{1}'$ is discontinuous at $\mu=0$ within this limit, with a quantized
discontinuity $-\left(\ell/4\right)\left[F'_{1}\left(0^{+}\right)-F'_{1}\left(0^{-}\right)\right]=\left(\ell/2\right)/96\pi$
that matches the coefficient $\beta$  of the \textit{gravitational
pseudo Chern-Simons} term of \cite{PhysRevB.98.064503}. As anticipated
in \cite{PhysRevB.98.064503}, the coefficient $c$ remains quantized
away from the relativistic limit, while $F_{1}'$ does not. Taking
the relativistic limit of the dimensionful coefficients \eqref{eq:c3},
one finds $F_{6}=0$, while $F_{4}=-\Delta_{0}^{2}F_{5}\neq0$ describe
a relativistic term which is second order in torsion, and was not
written explicitly in \cite{PhysRevB.98.064503,hughes2013torsional}. 

Finally, we note that our perturbative computation of the gCS term is analogous
to the computations of \cite{goni1986massless,van1986topological,vuorio1986parity,vuorio1986parityErr,kurkov2018gravitational}
for relativistic fermions, and reduces to these as $\kappa\rightarrow0$.


\twocolumngrid

\bibliographystyle{apsrev4-1}
\input{ManuscriptRefsV2.bbl}

\end{document}

%% file: ManuscriptRefsV2.bbl
%